\documentclass{aa}  
\usepackage{newtxtext,newtxmath}
\usepackage[T1]{fontenc}

\DeclareRobustCommand{\VAN}[3]{#2}
\let\VANthebibliography\thebibliography
\def\thebibliography{\DeclareRobustCommand{\VAN}[3]{##3}\VANthebibliography}

\usepackage[final]{changes}

\usepackage{hyperref}
\usepackage{xcolor}
\usepackage{amsmath}
\usepackage{natbib}
\usepackage{graphicx}
\usepackage{appendix}
\usepackage{float}

\usepackage{subfigure}
\usepackage{epstopdf}
\usepackage{amssymb}
\usepackage{amsmath}

\newcommand{\RNum}[1]{\uppercase\expandafter{\romannumeral #1\relax}}
\usepackage{color}
\definecolor{darkgreen}{rgb}{0.0,0.5,0.0}
\usepackage{array}
\usepackage{booktabs}
\usepackage{multirow}

\begin{document}

   \title{Galaxy-galaxy lensing in the VOICE deep survey}


   \author{Ruibiao Luo\inst{\ref{inst1},\ref{inst2}},
          Liping Fu\inst{\ref{inst2}},
          Wentao Luo\inst{\ref{inst3},\ref{inst4}},
          Nicola R. Napolitano\inst{\ref{inst1}},
          Linghua Xie\inst{\ref{inst1}},
          Mario Radovich\inst{\ref{inst5}},
          Jing Liu\inst{\ref{inst2}},
          Rui Li\inst{\ref{inst1}},
          Valeria Amaro\inst{\ref{inst2}},
          Zhu Chen\inst{\ref{inst2}},
          Dezi Liu\inst{\ref{inst6}},
          Zuhui Fan\inst{\ref{inst6}}, 
          Giovanni Covone\inst{\ref{inst7},\ref{inst8},\ref{inst9}},
          \and 
          Mattia Vaccari\inst{\ref{inst10},\ref{inst11},\ref{inst12}},         
          }    
   \institute{School of Physics and Astronomy, Sun Yat-sen University Zhuhai Campus, 2 Daxue Road, Tangjia, Zhuhai, Guangdong 519082, PR China\\
                \email{napolitano@mail.sysu.edu.cn}\label{inst1}
         \and
            Shanghai Key Lab for Astrophysics, Shanghai Normal University, Shanghai 200234, PR China\\
            \email{fuliping@shnu.edu.cn}\label{inst2}
         \and
            CAS Key Laboratory for Research in Galaxies and Cosmology, University of Science and Technology of China, Hefei, Anhui 230026, PR China\\
            \email{wtluo@ustc.edu.cn}\label{inst3}
         \and
            Department of Astronomy, School of Physical Sciences, University of Science and Technology of China, Hefei, Anhui 230026, PR China\label{inst4}
         \and
            INAF - Osservatorio Astronomico di Padova, via dell'Osservatorio 5, I-35122 Padova, Italy\label{inst5}        
         \and
            South-Western Institute for Astronomy Research, Yunnan University, Kunming 650500, Yunnan, PR China\label{inst6}    
         \and
            Dipartimento di Fisica 'E. Pancini', Universita degli Studi Federio \RNum{2}, Napoli I-80126, Italy\label{inst7}        
         \and
            INFN, Sezione di Napoli, Napoli I-80126, Italy\label{inst8}        
         \and
            INAF – Osservatorio Astronomico di Capodimonte, Salita Moiariello 16, Napoli I-80131, Italy\label{inst9}
         \and
         	Inter-university Institute for Data Intensive Astronomy, Department of Astronomy, University of Cape Town, 7701 Rondebosch, Cape Town, South Africa\label{inst10} 
         \and
			Inter-university Institute for Data Intensive Astronomy, Department of Physics and Astronomy, University of the Western Cape, 7535 Bellville, Cape Town, South Africa\label{inst11}         
         \and
         	INAF - Istituto di Radioastronomia, via Globetti 101, 40129 Bologna, Italy\label{inst12}
         }

   \date{Received xxx / Accepted xxx}
   
 \abstract{The multi-band photometry of the VOICE imaging data, overlapping with 4.9 deg$^2$ of the Chandra Deep Field South (CDFS) area, enables both shape measurement and photometric redshift estimation to be the two essential quantities for weak lensing analysis. The depth of $mag_{AB}$ is up to 26.1 (5$\sigma$ limiting) in $r$-band. We estimate the Excess Surface Density (ESD; $\Delta\Sigma$) based on galaxy-galaxy measurements around galaxies at lower redshift (0.10<$z_l$<0.35) while we select the background sources to be at higher redshift ranging from 0.3 to 1.5. The foreground galaxies are divided into two major categories according to their colour (blue/red), each of which has been further divided into high/low stellar mass bins. Then the halo masses of the samples are estimated by modelling the signals, and the posterior of the parameters are samples via Mote Carlo Markov Chain (MCMC) process. We compare our results with the existing Stellar-to-Halo Mass Relation (SHMR) and find that the blue low stellar mass bin (median $M_*=10^{8.31}M_\odot$) deviates from the SHMR relation whereas all other three samples agrees well with empirical curves. We interpret this discrepancy as the effect of a low star formation efficiency of the low-mass blue dwarf galaxy population dominated in the VOICE-CDFS area. 
 }

   \authorrunning{Ruibiao Luo et al}
   \maketitle

%

\section{Introduction}
A major challenge of galaxy formation is to understand the co-evolution processes of gas, stellar, and dark matter in galaxies as a function of their properties, such as mass and colour \citep{Wechsler2018}.
Theoretical studies (see e.g. \citealt{White1978,Fukugita1998,Faucher2011,Somerville2015}) suggest that the physical progresses of galaxy formation are driven by the properties of their dark matter haloes, in particular their mass. Hydrodynamical simulations have recently reached a sufficient accuracy to study the effect of the stellar feedback and other strong mechanisms like active galactic nucleus (AGN) and supernovae feedback to rather small scales (Illustris, \citealt{Vogelsberger2014}; EAGLE, \citealt{Schaye2015}), also allowing us to study the effect of gas and stellar processes on the final dark matter distribution \citep{White1978,Blumenthal1986,dave2012}. Ultimately, we expect these simulations to finally bridging the baryonic and dark matter properties \citep{yang2006} and solve the so called galaxy-halo connection \citep{Wechsler2018}.

Observationally speaking, abundance matching \citep{Conroy2009,Behroozi2010,Moster2013}, such as the the relation between the stellar, $M_*$, and dark matter (DM) mass, $M_{\rm DM}$, in halos obtained by matching the the observed galaxy luminosity function and the predicted halo mass function from simulations (see e.g. \citealt{tinker2005,Vale2006,Conroy2006}), is one of the primary semi-empirical test of the existence of such a connection. Other popular methods are the Halo Occupation Distribution (HOD; \citealt{Berlind2002,Kravtsov2004,Zheng2005,zu2016}) that populates dark matter haloes with galaxies to reproduce galaxy clustering \citep{Jing1998,Seljak2000,Peacock2000,Cooray2002} as a function of luminosity over a wide redshift range \citep{Vale2004,Conroy2006}. Furthermore, there are more complex methods developed with the HOD approach, such as the conditional luminosity function (CLF; \citealt{yang2003}) and conditional stellar mass function (CMF; \citealt{Moster2010}).
These methods constitute hybrid approaches based on statistical relations between observed galaxies and simulated halos, and, as such, they are strongly model-dependent. 

On the other hand, to fully test the theoretical expectation, one should be able to have a direct measurement of both the stellar and the dark component of galaxies to construct a $M_*-M_{\rm DM}$ relation.
One possibility is to use dynamical-based methods to obtain the total mass in galaxies (see e.g. \citealt{Blumenthal1986,Zaritsky1994,McKay2002,Prada2003}).
Another possibility is provided by the gravitational lensing. This is a powerful technique to infer the galaxy masses at different scales. In the case of strong lensing, arcs or multiple images of background 'source' galaxies allow to efficiently constrain the total mass in the central regions of foreground 'lens' systems \citep{Kochanek1995,Treu2010}. In case of weak lensing (WL hereafter), the effect of the weak distortion over large statistical sample of background galaxies can be used to infer the total mass density out to very large distances for an ensemble of foreground lens systems \citep{Brainerd1996,Bartelmann2001,Munshi2008,Hoekstra2008}.  
In this latter case, specifically, 
we refer to galaxy-galaxy lensing, to distinguish it from other forms of weak lensing from larger distribution of matter at cluster (see e.g. \citealt{Natarajan1997,Geiger1999}) or cosmic scales (see e.g. \citealt{Mandelbaum2013,Kwan2017}).

In the last few decades, there have been large progresses inweak gravitational lensing studies from wide-field and deep sky surveys. These have provided high-quality photometric images for weak lensing,
such as the Sloan Digital Sky Survey (SDSS, \citealt{york2000}; \citealt{Guzik2002}; \citealt{cacciato2009,Cacciato2013}; \citealt{luo2018}), the Canada-France-Hawaii Telescope Lensing Survey (CFHTLenS, \citealt{heymans2012}; \citealt{Kilbinger2013}; \citealt{Fu2014}; \citealt{hudson2015}), Dark Energy Survey (DES, \citealt{descollaboration2005}; \citealt{DEScollaboration2016}; \citealt{clampitt2017}; \citealt{Abbott2018}), Kilo-Degree Survey (KiDS, \citealt{Kuijken2015}; \citealt{viola2015}; \citealt{Uitert2018}; \citealt{Dvornik2020}), Hyper-Suprime-Cam survey (HSC, \citealt{Aihara2018}; \citealt{Wang2021}), etc. Because of the variety of astrophysical answers the WL can provide about DM, this has become the main science driver for most of future larger survey projects. Future space-based surveys will be provided by the missions of $Euclid$ \citep{refregier2010} and $Roman$ \citep{Spergel2015}, and Chinese Space Station Optical Survey (CSS$\verb|-|$OS, \citealt{zhan2011,zhan2018,Gong2019}). About the future ground-based survey, there is the Legacy Survey of Space and Time (LSST, \citealt{LSST2009}) for next decade. 

In this paper, we are focussing on the CDFS region of VOICE survey that is the VST Optical Imaging of the CDFS and ES1 Fields survey \citep{vaccari2016}, and we estimate the two-dimensional Excess Surface Density (ESD) of galaxy-galaxy lensing from the measurements of tangential shape signals of sources from the shear catalogue in VOICE$\verb|-|$CDFS which are presented in \cite{fu2018} (hereafter $F18$). We apply Markov Chain Monte Carlo (hereafter MCMC) method to build a halo model that constrain the halo parameters of foreground galaxies, and finally directly derive the relation between the stellar and halo mass for central and satellite galaxies \citep{yang2006} in the VOICE$\verb|-|$CDFS region.

The structure of this paper is described as follows. In Section 2, we present the dataset from the VOICE survey. In Section 3, we illustrate the shear catalogue from the background source sample and the selection of foreground (lens) samples based on the photometric catalogue of the galaxies in VOICE. In Section 4 we introduce the galaxy-galaxy lensing estimator, while in Section 5 we present the weak lensing model we adopt to estimate the halo parameters. The ESD measurements and the model results are presented in Section 6, together with a comparison of the ESD results obtained using the DES-Y1 shear catalogue overlapping with VOICE on the CDFS area. In Section 7 we finally discuss the results and draw some conclusions.


\section{VOICE Survey and Shear catalogue}
The VOICE Survey is a Guaranteed Time of Observation (GTO) survey carried out with the European Southern Observatory (ESO) VLT Survey Telescope (VST; \citealt{capaccioli2011}) on Cerro Paranal in Chile. VOICE observations have been carried out from October 2011 to 2015 to obtain deep optical imaging of two patches of the sky, each of about 5 deg$^2$, centred on the Chandra Deep Field South (CDFS) and on the European Large Area ISO Survey South$\verb|-|$1 (ES1). The two areas are dubbed VOICE$\verb|-|$CDFS ($\verb|RA|$=$03^h32^m30^s$, $\verb|DEC|$=$-27^{\circ}48'30''$) and VOICE$\verb|-|$ES1 ($\verb|RA|$=$00^h34^m45^s$, $\verb|DEC|$=$-43^{\circ}28'00''$), respectively. 
    
These two area have been targeted in the past from different projects and in different wavelength ranges, including ultraviolet (UV) from GALEX \citep{Martin2005}, near-infrared (NIR) band from VISTA/VIDEO \citep{jarvis2013}, mid-infrared (MIR) band from Spitzer$\verb|-|$Warm/SERVS \citep{Mauduit2012}, far-infrared (FIR) from Herschel/HerMES \citep{Oliver2012}, infrared (IR) from Spitzer$\verb|-|$Cold/SWIRE \citep{Lonsdale2003}, and radio band in ATCA/ATLAS \citep{Franzen2015}. VOICE was designed to provide deep, high-quality observations in $ugri$ bands on VOICE$\verb|-|$CDFS field and $u$-band on VOICE$\verb|-|$ES1 field. 

In this paper, we use  $r$-band data of the 4.9 deg$^2$ area of VOICE$\verb|-|$CDFS. This is composed of four pointings (CDFS$\verb|-|$1/2/3/4), observed with the $\verb|OmegaCAM|$ \citep{Kuijken2011} consists of 32 detectors with 2048$\times$4096 pixels and a scale of 0.21 arcsec/pixels, using the same tiling strategy adopted for weak lensing observations in KiDS \citep{Kuijken2019}. There are more than $100$ $r$-band exposures in each of the four tiles, making this the deepest band available in VOICE. For the four different area, the cumulative exposure time is in the range of $15.30$ to $20.90\ h$. Just like in KiDS, the observations consisted in five continuous exposures every epoch by repeating a diagonal pattern to cover the detector gaps between charge-coupled devices (CCDs). The VOICE data we used in galaxy-galaxy lensing study is based on the shear catalogue from F18, where the galaxy shapes have been measured by $\verb|LensFit|$ \citep{miller2013}. The VOICE shear catalogue has been derived by the $r$-band stacked images, as a product of an analysis pipeline including image co-adding, star and badpixel masking, object detection,  PSF fitting, shape measurements, etc (see F18). 
    
The final catalogue of objects classified as galaxies, is made of 583131 objects (see F18 for details). For these galaxies, the measurements of photometric redshifts (photo-z, hereafter) was based on the data of the optical observations in $u$, $g$, $r$, $i$ from VOICE together with the NIR observations in $Z$, $Y$, $J$, $H$, $K_s$ from the VIDEO survey \citep{jarvis2013}, using the $\verb|BPZ|$ software \citep{benitez2011}. We will refer to this as the {\it photometric catalogue}, in the following. 
    
Finally, the {\it shear catalogue} has been obtained using $\verb|LensFit|$ \citep{miller2013} as the shape measurement algorithm for $\verb|OmegaCAM|$ images. In particular, the weak lensing shear measurements are based on $r$ band images with $\leq$0.9 arcsec seeing in VOICE survey. 
    
The shear catalogue of VOICE$\verb|-|$CDFS contains the data of 310985 galaxies corresponding to an effective weighted galaxy number density about 16.35 gal arcmin$^{-2}$, which, as comparison, is about twice the density in KiDS survey. The limiting AB magnitude for a point source in 2 arcsec aperture is 26.1 mag in $r$-band. We refer the interested reader to F18 for all further details about the data reduction and the validation of the shear and photo-z catalogues.


\section{Galaxy Sample}
\label{sec:gal_sample}
The photometric galaxy catalogue and the shear catalogue have different purposes. The former provides the photometric information of all galaxies identified in the CDFS area in VOICE. These are all galaxies that can be used as potential lenses, at different redshifts, in our galaxy-galaxy lensing estimates. The photo-z derived from $\verb|BPZ|$ has accepted systematics and F18 presents there is a good agreement between photo-z with spectroscopic redshift (spec-z, hereafter)
for matched galaxies. The shear catalogue, instead, represents the list of galaxies for which we have been able to measure the apparent distortion due to the weak lensing effect. As such, this is the catalogue where we need to select the background sources in the surrounding area of each foreground lens chosen from the photometric catalogue.

\subsection{Photo-z and stellar masses}
\label{sec:stell_mass}
\begin{figure}
	\centering
		\includegraphics[width=8.5cm]{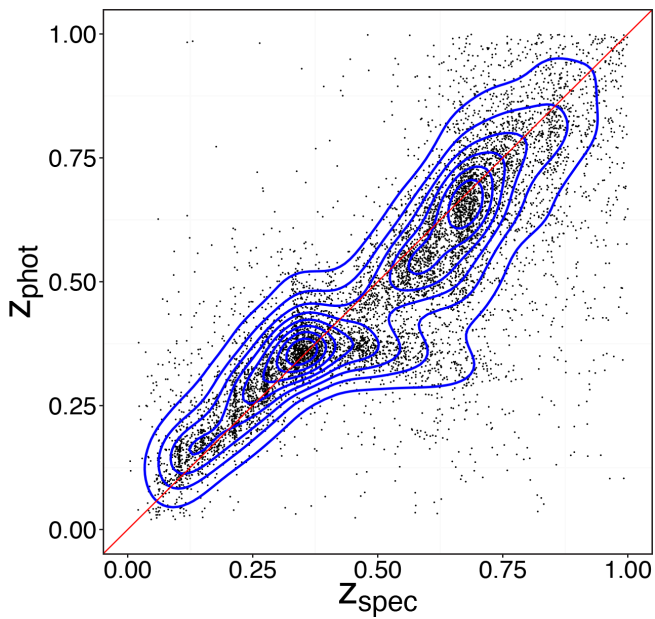}
		\caption{Comparison of photo-z ($Z_{phot}$) versus spec-z ($Z_{spec}$) for the matched galaxies (black points) sample \citep{fu2018}. The contours present the galaxy number density. The red line is the one-to-one relation.
		}
		\label{fig:photz_vs_specz}
\end{figure}
As discussed in F18, the photo-z estimation in the VOICE study is the peak value of the probability density function, and the photo-z data was derived using the $\verb|BPZ|$ software. We checked the comparison of the measurements of photo-z with the corresponding spec-z \citep{Vaccari2010,vaccari2016} for the matched 23638 galaxies in Fig.~\ref{fig:photz_vs_specz}, and It shows good agreements to spec-z for corresponding photo-z on the whole. F18 presents the median value of the difference between photo-z and spec-z: $\delta z=(Z_{phot}-Z_{spec})/(1+Z_{spec})=0.008$, with median absolute deviation $\sigma=0.06$. For the photometric catalogue and shear catalogue, we consider the uncertainties of sources from $\verb|BPZ|$ are good enough to support the photo-z used for estimating galaxy-galaxy lensing signals.

Galaxy masses are derived usind a standard spectral energy distribution (SED) fitting software, {\tt Le-Phare}\footnote{\url{http://www.cfht.hawaii.edu/~arnouts/lephare.html}}\citep{Arnouts1999,ilbert2006}. 
Since we want to ultimately study the halo properties of the lens sample, and relate these to their stellar mass properties, we have used the photometric galaxy multi-band (optical plus NIR) catalogue to estimate the stellar masses. Here {\tt Le-Phare} is fed with the full 9-band photometry from the VOICE galaxy catalogue to produce, as output, the best stellar population parameters, including the age, metallicity, star formation rate and the stellar mass.  
The stellar population synthesis (SPS) models \citep{Conroy2009} we have adopted to match the multi-band photometry are stellar templates from \cite{Bruzual2003} with a \cite{Chabrier2003} initial mass function (IMF) and an exponential decaying star formation history. 

For the {\tt Le-Phare} run, we have used a broad set of models with different metallicities ($0.005 \leq Z$/${Z_{\odot}} \leq 2.5$) and ages ($age \leq age_{\rm max}$), with the maximum age, $age_{\rm max}$, set by the age of the Universe at the redshift of the galaxy, with a maximum value at z = 0 of $13.5\, \rm Gyr$.  We also considered internal extinction using the \cite{Calzetti1994} models. Finally, to reduce the degeneracies between the redshift and galaxy colours, we have fixed the galaxy redshift to the VOICE catalogue photo-z.

\subsection{Lens sample}
\label{sec:lens_sample}
\begin{figure}
	\centering
		\resizebox{\hsize}{!}{\includegraphics{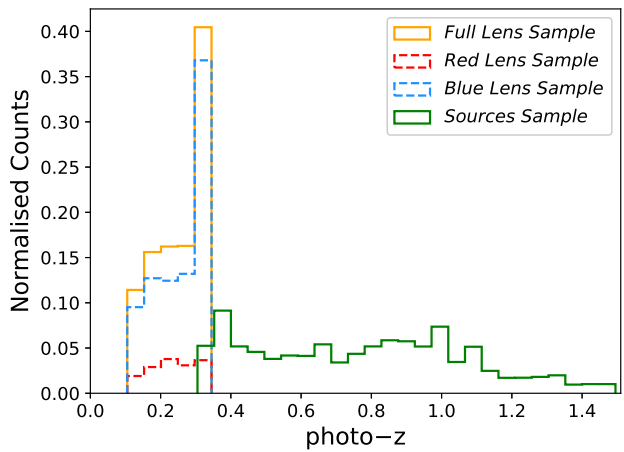}}
		\caption{Normalised distribution of photo-z (BPZ) for the galaxies of Full Lens Sample (FLS, orange histogram) and Sources Sample (green histogram). The distributions of Red/Blue Lens Sample (red/blue dashed histogram) are normalised to FLS. The redshift range of FLS is $0.1<z_l<0.35$ and of Sources Sample is $0.3<z_s<1.5$.
		}
		\label{fig:lens_source_z}
\end{figure}

As mentioned in Sect.~\ref{sec:gal_sample}, the lens sample is based on the photometric galaxy catalogue, regardless any shear has been measured for them.
Among these, we have selected galaxies in the range of $0.1< z_l(\verb|BPZ|)<0.35$, for which all $gri$-bands magnitudes are available (hereafter 'Full Lens sample', or FLS in short). This catalogue is made of 46188 galaxies, containing positions and photo-z for each of them. 
The distribution of photo-z of this FLS, shown in Fig.~\ref{fig:lens_source_z}, has a median of 0.29 presents that the galaxies are dominated FLS in the redshift bin from 0.30 to 0.35. 
The choice of this specific redshift range for the FLS is made to maximise the number of foreground galaxies to guarantee a galaxy density minimising the statistical errors over the two-dimensional lensing signal which is represented in Sect.~\ref{sec:ESD}. 

The mean luminosity of the FLS is $M_r\sim-18.06$ with a scatter of $\sigma(M_r)\sim 1.61$ mag, while the averaged logarithmic stellar mass is $\log M_*/M_\odot=8.56$, with a scatter of $\sigma(\log M_*/M_\odot)=0.96$. This rather low mean value and large scatter imply that a large portion of the lens systems have low mass. 
Indeed, the stellar masses are distributed in the range $10^6-10^{12} M_\odot$, such as they cover a very large mass range going from dwarf galaxies to giant ellipticals. As we seek to obtain mean dark halo properties of the lens sample, which are a strong function of the stellar mass (e.g. \citealt{Moster2013}), averaging the mass profiles in this wide range of masses is poorly meaningful. Hence, we have decided to bin galaxies in stellar mass. 

Furthermore, to study the halo properties as a function of galaxy types (e.g.
\citealt{Mandelbaum2006b,hudson2015}), we roughly separate the passive red galaxies from some active bluer systems, we further bin them in colour. 

\begin{figure}
	\centering
		\resizebox{\hsize}{!}{\includegraphics{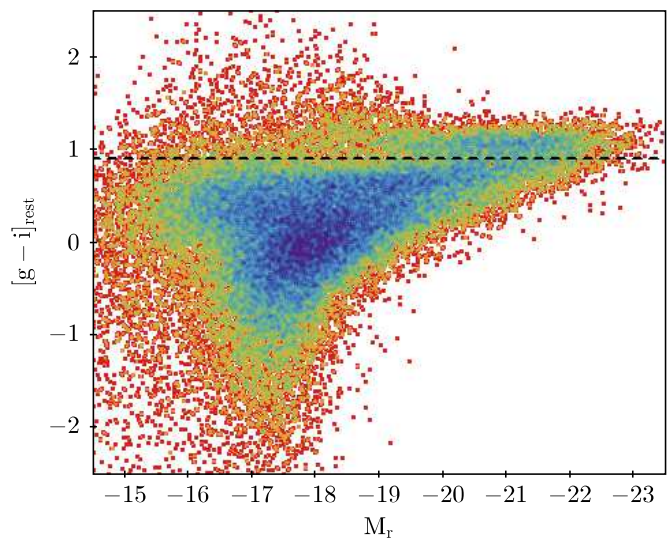}}
		\caption{Distribution of colour [g-i] (rest-frame) versus $r$-band absolutely magnitude for the FLS. The colour of points (from red to blue) encodes increasing galaxy number densities. The black dashed line is the criterion of $[g-i]_{rest}=0.9$ to separate red/blue galaxies. There are two sequences of galaxies in $[g-i]_{rest}>0.9$ and $[g-i]_{rest}<0.9$ that are considered as the galaxies of Red Lens and Blue Lens, respectively.}
		\label{fig:color_Mr}
\end{figure}

The distribution of galaxy colours as a function of the $r-$band rest-frame magnitude is shown in Fig.~\ref{fig:color_Mr}. Here we can clearly identify a red sequence at the rest frame $[g-i]_{rest}>0.9$, for $M_r<-19$. Hence we have separate FLS into the Red and Blue lenses as the ones above and below the dashed line in Fig.~\ref{fig:color_Mr}, respectively. We have tried to use other classification of red and blue galaxies like $[u-g]_{rest}$, $[g-r]_{rest}$ \citep{bell2003}, $[u-r]_{rest}$ \citep{Baldry2004},
but these other methods hardly catch the obvious bimodal distribution in the colour magnitude diagram for the classification of FLS as seen in Fig. \ref{fig:color_Mr}. 
Fig. \ref{fig:lens_source_z} shows the distributions of photometric redshifts of the Red and Blue lenses with that the means are 0.24 and 0.26, respectively.  

\begin{figure}
	\centering
		\resizebox{\hsize}{!}{\includegraphics{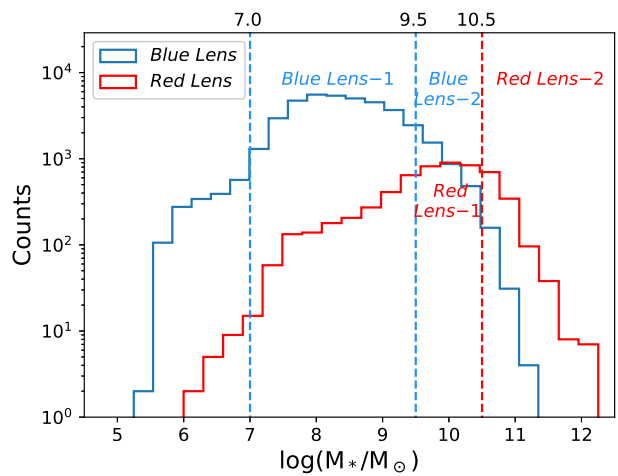}}
		\caption{Distribution of logarithmic stellar mass from Blue Lens (blue histogram) and Red Lens (red histogram). The Blue Lens-1 $\&$ 2 are from the logarithmic stellar mass bins of [7.0, 9.5] $\&$ [9.5, 10.5] in Blue Lens. The Red Lens-1 $\&$ 2 are from the logarithmic stellar mass bins of [9.5, 10.5] $\&$ $>$10.5 in Red Lens.}
		\label{fig:red_blue_mass}
\end{figure}
	
In Fig.~\ref{fig:red_blue_mass} we use the colour separation above to display the mass distribution of the two colour classes. This shows a clear bimodal distribution in $\log M_*$ for the Red and Blue Lens. The Red Lens sample contains 5822 galaxies and the Blue Lens sample 40366 with median $\log M_*/M_\odot$ of $9.88$ and $8.37$, respectively.
Looking at the distributions of the stellar masses in Fig.~\ref{fig:red_blue_mass}, we can see that the most massive bin, such as $\log M_*/M_\odot>10.5$ is mainly occupied by the Red lenses, while in the bin $9.5<\log M_*/M_\odot<10.5$ there is a mix of Blue and Red lenses, although the former are dominant in absolute number. In the lower mass range, 7.0<$\log M_*/M_\odot$<9.5, the Blue lenses reach their peak at $\log M_*/M_\odot\sim8.5$, while they start to become incomplete at lower masses. The Red lenses, though, are too little in the same mass bin, $8.5<\log M_*/M_\odot<9.5$, to produce a significant lensing signal. Hence to obtain a significant colour/mass separation of the FLS, we have defined the following samples:   
\begin{enumerate}\setcounter{enumi}{0}
    \item Blue Lens$\verb|-|$1 and $\verb|-|$2: $10^{7.0}<M_*/M_\odot<10^{9.5}$, $10^{9.5}<M_*/M_\odot<10^{10.5}$, respectively;
    \item Red Lens$\verb|-|$1 and $\verb|-|$2: $10^{9.5}<M_*/M_\odot<10^{10.5}$, $M_*/M_\odot>10^{10.5}$), respectively.
\end{enumerate}

In Table~\ref{Tab:01} we summarise the galaxy number, the median of photo-z and logarithmic stellar mass for the four lens sub-samples. Hereafter, we will make use of the median values of these parameters to characterise the four lens samples.

\begin{table*}\footnotesize
\renewcommand\arraystretch{1.6}
	\centering
	\caption{Statistics of Red Lens-1/2, Blue Lens-1/2, and Full Lens with the mass bin range, numbers, the median photo-z and median logarithmic stellar mass}
	\label{Tab:01}
	\begin{tabular}{lcccc}
		\hline\hline
		Lens Sample & $\log{(M_*/M_\odot)}$ Range  & Number & $ z_l({\verb|BPZ|})$ & $\log{(M_*/M_\odot)}$\\
		\hline
		{$Blue\quad Lens-1$} & 7.0$-$9.5 & 34770 & 0.30 & 8.31 \\

		{$Blue\quad Lens-2$} & $9.5-10.5$ & 3703 & 0.29 & $9.79$ \\
		\hline
		{$Red\quad Lens-1$} & $9.5-10.5$ & 2834 & 0.27 & $10.01$ \\

		{$Red\quad Lens-2$} & $>10.5$ & 1101 & 0.28 & $10.73$ \\
		\hline
		{$Full\quad Lens$} & full & 46188 & 0.29 & $8.49$  \\
		\hline
	\end{tabular}
\end{table*}

\subsection{Source sample and lens-source pairs}
\label{sec:lens_source_pair} 
As background galaxy sample (the sources), we select galaxies from the shear catalogue of VOICE$\verb|-|$CDFS with photo-z in the range of $0.3<z_s(\verb|BPZ|)<1.5$. The distribution of photo-zs is shown in Fig.~\ref{fig:lens_source_z}. 

The lens-source pairs are then selected using the condition that $\Delta z_p = z_s-z_l > 0.2$. This criterion has been adopted to take into account the errors on the photometric redshifts and avoid confusion between background and foreground objects if too close in redshift. According to F18, the typical photo-z errors are $\sim0.06$ for $z<0.83$ and $\sim0.1$ for $z>0.83$, hence the use of $\Delta z_p>0.2$ allows us to separate foreground from background with $\sim2\sigma$ significance.


\section{Galaxy-galaxy lensing estimator}
\label{sec:gg_theory}
\subsection{Tangential shear}
\label{sec:tang_shear}

Galaxy-galaxy lensing signal estimator is based on the measurement of tangential shear $\gamma_t$ from background sources, around the foreground lenses. As mentioned before, galaxy shapes have been measured in F18 using $\verb|LensFit|$, where galaxy ellipticities  $\gamma_1$ and $\gamma_2$ have been derived from $\verb|OmegaCAM|$ images with an accuracy to $\leq$1$\%$. 
These measurements are used to estimate the tangential component $\gamma_t$ and cross component $\gamma_\times$ of shear signals of background sources around a lens galaxy for the $j$-th lens-source pair according to the equations
\begin{equation}
	    \left[ 
        \begin{array}{c} \gamma_{t,j} \\ \gamma_{\times,j}\\ \end{array} 
	    \right]
	    =\left[ 
        \begin{array}{cc} -\cos{(2\phi_j)} & -\sin{(2\phi_j)} \\ \sin{(2\phi_j)} & -\cos{(2\phi_j)}\\ \end{array} 
	    \right]
	    \left[ 
        \begin{array}{c} \gamma_{1,j} \\ \gamma_{2,j}\\ \end{array} 
	    \right],
\end{equation}
where $\phi_j$ is the angle between the separation vector of the $j$-th lens-source pair with the horizontal axis in Cartesian coordinate system centred on each object of lens. 
In weak lensing, the weak distortion of the intrinsic shape due to the warped space-time caused by the lenses of each independent background source is too small to be detected.
Hence, in order to detect the shear signals, we need to average over large numbers of lens-source pairs to finally measure, in particular, the tangential component of the shear.
This allows us to derive the signal around a lens sample in angular bins $\theta$ \citep{mandelbaum2005,luo2018},
\begin{equation}
	    \overline{\gamma}_t(\theta)=\frac{1}{2\overline{\mathcal{R}}}\frac{\Sigma_j w'_j\gamma_{t,j}}{\Sigma_j w'_j},
\end{equation}
where $\overline{\mathcal{R}}$ means the responsivity of source galaxies derived by the equations (5) to (7) in \cite{Jarvis2003}, and here $w'_j$ is the weight come from $\verb|LensFit|$ for the $j$-th lens-source pair.

This quantity is used next to derive a proxy of the mass density as a function of the angular distance from the common centre of the lens sample adopted to measure it.

\subsection{Excess Surface Density (ESD)}
\label{sec:ESD}
The Excess Surface Density (ESD; $\Delta\Sigma$) is defined as the discrepancy between $\overline{\Sigma}(\leq R)$ with $\overline{\Sigma}(R)$ that are the averaged projected surface mass densities inside of radius $R$ and at radius $R$, 
\begin{equation}
    \Delta\Sigma(R; z_l)  = \overline{\Sigma}(\leq R)-\overline{\Sigma}(R).
    \label{eq:esd}
\end{equation}

There is a connection between the ESD with the tangential shear from background sources. Indeed the $\Delta\Sigma$ can be written as \citep{hudson2015}:
\begin{equation}
    \begin{aligned}
    \Delta\Sigma(R; z_l) & = \Sigma_{\rm crit}(z_l,z_s)\overline{\gamma}_t (R; z_l,z_s)\\
    & = \frac{\Sigma_j[w_j\gamma_{t,j}(R; z_l,z_s)/\Sigma^{-1}_{{\rm crit,}j}(z_l,z_s)]}{\Sigma_j w_j},
    \label{eq:esd_shear_t}
    \end{aligned}
\end{equation}
where the critical surface density $\Sigma_{\rm crit}$ is defined as 
\begin{equation}
	\Sigma_{\rm crit}(z_l,z_s)=\frac{c^2}{4\pi G}\frac{D_s}{D_l D_{ls}},
\end{equation}
where $D_l$, $D_s$,  and $D_{ls}$ are the angular diameter distance of the lens, background source, and that between two the objects, respectively. 

In this equation, pairs are weighted by the $\Sigma^{-2}_{{\rm crit,}j}(z_l,z_s)$, such that we write the weights $w_j$, for the $j$-th lens-source pair, as
\begin{equation}
    	w_j=w'_j\Sigma^{-2}_{{\rm crit,}j}(z_l,z_s).
    \label{eq:weight}
\end{equation}
    
Hence, Eq.~(\ref{eq:esd_shear_t}) states that we can estimate the ESDs directly from the mean tangential shear signal of sources around the lenses in the different bins of projected separation $R$. 

However, we need to correct the shear for possible biases in the shear measurements by $\verb|LensFit|$. The shear calibration \citep{liu2018}, brings a multiplicative, $m$, and an additive, $c$, bias into the shear estimation, that allow us to convert the observed shear into a 'true' signals as
\begin{equation}
    	\gamma^{\rm obs}_a = (1+m_a)\gamma^{\rm true}_a+c_a,
    	\label{eq:gamma_obs}
\end{equation}
where $a$ presents two components ($a$=1, 2) of galaxy ellipticities. 
This calibration can be applied to our averaged ESD measurement as above to obtain a corrected mean ESD measurement. This is a function of lens redshift and writes as follows:
\begin{equation}
    \begin{aligned}
	    \Delta\Sigma^{\rm lens}(R)=&\frac{1}{2\overline{\mathcal{R}}}\frac{\Sigma_j w_j[-(\gamma_{1,j}-c_{1,j})\cos{2R_j}]\Sigma_{\rm crit}}{\Sigma_j w_j (1+m_{1,j})}\\
	    &+\frac{1}{2\overline{\mathcal{R}}}\frac{\Sigma_j w_j[-(\gamma_{2,j}-c_{2,j})\sin{2R_j}]\Sigma_{\rm crit}}{\Sigma_j w_j (1+m_{2,j})},
    \end{aligned}
\end{equation}
where $c_1$ and $c_2$ are the additive biases and $m_1$ and $m_2$ are the multiplicative biases, obtained as discussed in $F18$. The estimated values of $c_1$ and $c_2$ are $\sim$8$\times10^{-4}$ and $\sim$3$\times10^{-5}$ for $\gamma_1$ and $\gamma_2$, respectively (see $F18$). Being these values $<<1$, the produce produce almost no effect on the shear measurements. On the other hand, the multiplicative biases $m_1$ and $m_2$ are quite uniformly distributed in the range (-0.494,0.678) and (-0.362, 0.696), respectively, hence they have to taken into account. 

Finally, to derive an unbiased ESD estimator, we need to subtract the tangential shear measurements around random points that ought not to have net lensing signal. This writes as:
\begin{equation}
    	\Delta\Sigma(R)=\Delta\Sigma^{\rm lens}(R)-\Delta\Sigma^{\rm rand}(R).
\end{equation}
According to the random test in Sect. \ref{sec:systematics}, however, the ESD measurements from tangential shear signals of the random sample, $\Delta\Sigma^{\rm rand}(R)$, is generally consistent with zero. Here we would consider the noise, $\Delta\Sigma^{\rm rand}(R)$ of 100 times the number of random points corresponding to our samples, is subtracted in our final ESD measurements.  

\begin{figure}
    \centering
    	\resizebox{\hsize}{!}{\includegraphics{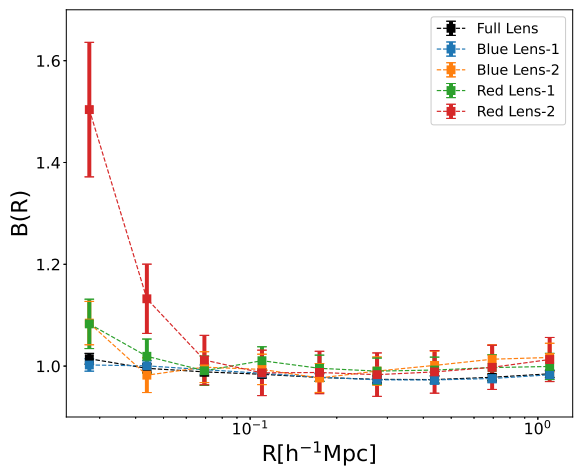}}
    	\caption{Boost factors B(R) for the background sources sample around the galaxies of FLS (black), Blue Lens-1/2 (blue/orange), Red Lens-1/2 (green/red) in 9 radial bins from approximately 0.03 to 1.2 $h^{-1}Mpc$. }
    	\label{fig:bfc}
\end{figure}

\subsection{Boost Factor}
Despite we have payed attention to avoid the overlap between the lens and source pairs by considering the $z_s-z_l$ separation in photo-z larger than 0.2 (see Sect. \ref{sec:lens_source_pair}), there can be still a fraction of background sources which are physically connected to the lenses, causing a scale-dependent bias of lensing signal due to galaxy clustering \citep{sheldon2004}. 
To correct for the effect of this correlation between lens with background sources, we apply a multiplicative boost factor $B(R)$. 
This is defined as the ratio between the weighted number of background galaxies per unit area around the lens and the ones around random points:
\begin{equation}
    	B(R)=\frac{n_{\rm lens}(R)}{n_{\rm rand}(R)}=\frac{N_{\rm lens}/\Sigma_{i,j}w_{i,j}}{N_{\rm rand}/\Sigma_{k,l}w_{k,l}},
\end{equation}
where $i,j$ and $k,l$ denote the sources found around the real lens and the random points, respectively, and $w_{i,j}$ or $w_{k,l}$ are the weight for the pair between each background source with one lens or a random point \citep{sheldon2004}.

In Fig.~\ref{fig:bfc} we show the $B(R)$ in radial bins from $\sim$0.03 to 1.2 $h^{-1}$Mpc for the different selected samples. In particular, we see that the $B(R)$ is close to one at all radii only for the FLS and the low mass Blue lens sample, while for all other samples it become significantly larger than one for $R<0.05$ $h^{-1}$Mpc, with the Red massive sample showing the largest factor at all radii, from 1.1 at $R>0.05$ $h^{-1}$Mpc to 1.6 for $R\sim0.03$ $h^{-1}$Mpc.
The similarity of the $B(R)$ between the FLS and the low mass Blue lenses comes mainly from the fact that the FLS is numerically dominated by the Blue Lens$\verb|-|$1 sample, which, because of the large statistics and the more sparse distribution in space (i.e. low mass blue galaxies tend to be less clustered than red massive galaxies, \citep{Zehavi2005}), has a lower chance to have an intrinsic excess of concentration. On the other end of the colour-mass selection, Red massive galaxies are known to cluster more \citep{Zehavi2005} as they are, for example, the dominant population in cluster of galaxies.     
 
The $B(R)$ of Blue Lens$\verb|-|$1 tend to 1 which is similar for FLS that presents the selection of lens-source pairs $\Delta z_p>0.2$ can clearly separate foreground lens and background sources out for the low stellar mass galaxies. The Blue Lens$\verb|-|$2 and Red Lens$\verb|-|$1 have a little bigger $B(R)$ tells there are increasing correlation between lens and sources. Especially for high stellar mass galaxies such as the Red Lens$\verb|-|$2, the boost factor reflects there is an obvious correlation effect we discussed before for lens-sources pairs. This allows us to compute an 'unbiased' $\Delta\Sigma$ by multiplying the $\Delta\Sigma$ measured as in Sect. \ref{sec:ESD} by the boost factor $B(R)$.

\subsection{Covariance Matrix}

\begin{figure}
    \centering
	    \resizebox{\hsize}{!}{\includegraphics{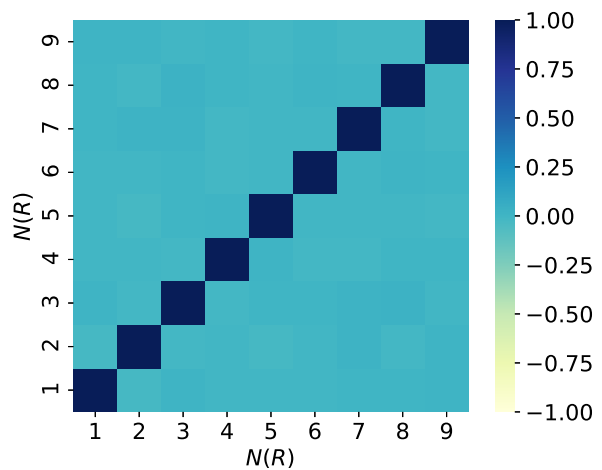}}
	    \caption{Correlation matrix is propagated from the estimated $\Delta\Sigma$ variance matrix of FLS in different radial bins according to the pipeline of bootstrapping. The label N(R) represents nine radial bins from approximately 0.03 to 1.2 $h^{-1}Mpc$. The colours from light yellow to deep blue mean the correlations of $\Delta\Sigma$ between radial bins are from weak to strong. }
	    \label{fig:CovarianceMatrix}
\end{figure}

To estimate the statistic errors on $\Delta\Sigma(R)$, we apply a bootstrap method to the covariance matrix of the $\Delta\Sigma(R)$. The dimensionless covariance matrix can be simply calculated by 
\begin{equation}
    	C_{i,j}=\frac{V_{i,j}}{\sqrt{V_{i,i}V_{j,j}}}
\end{equation}
where $V_{i.j}=\langle(\Delta\Sigma(R_i)-\langle\Delta\Sigma(R_i)\rangle)\cdot(\Delta\Sigma(R_j)-\langle\Delta\Sigma(R_j)\rangle)\rangle$, and it makes sense by changing corresponding index for $V_{i,i}$ and $V_{j,j}$. If $\Delta\Sigma(R)$ is subtracted by the random signal $\Delta\Sigma^{rand}(R)$, the ESD measurements will gain smaller covariance \citep{Singh2016}. 

We calculate the covariance matrix from the correlations between different $\Delta\Sigma(R)$, which are re-sampled by bootstrapping for $10^4$ times, around the foreground galaxies from FLS objects. In our cases, the Hartlap correction \citep{Hartlap2007} $(N_s-N_{points}-2)/(N_s-1)$ is negligible for covariance matrix cause it is very close to unity, where $N_s=10^4$ is the number of simulations and $N_{points}=9$ is the data points. In Fig.~\ref{fig:CovarianceMatrix}, the off-diagonal terms shows there are not highly correlation of $\Delta\Sigma(R)$ between different radial bins, but on the contrary, there are strong self-correlation of $\Delta\Sigma(R)$ that are reflected in the diagonal of the covariance matrix. This means that ESDs from different radial bins are mutually independent in the galaxy-galaxy lensing signal. Furthermore, we can calculate the reduced $\chi^2$ to qualify the goodness of the fit between model and data, using the equation
\begin{equation}
    	\chi^2_{d.o.f}= (data-model)^T C^{-1} (data-model)/d.o.f,
\end{equation}
where $C^{-1}$ is the inverse covariance matrix, and degree of freedom $d.o.f=N_{points}-N_{para}-1$, where $N_{para}$ are the number of model parameters. Table~\ref{Tab:02} also shows the $\chi^2_{d.o.f}$ for the model results correspond to lens samples.


\section{Weak-lensing Model}
\subsection{The Model}
\label{sec:halo_model}
The measurements of $\Delta\Sigma(R)$ can be derived from the galaxy-matter cross-correlation $\xi_{gm}(r)$ so that the galaxy-galaxy lensing  can give an effective estimator of the dark matter halo profile and galaxy environment in the area around the lens. The $\xi_{gm}(r)$ is the line-of-sight projection of galaxy-matter cross-correlation function, defined as \citep{luo2018}:
\begin{equation}
    	\xi_{gm}(r)=\langle\delta(x)_g\delta(x+r)_m\rangle,
\end{equation}
which relates the surface mass density to a corresponding lens galaxy. The ESD would be calculated by the Eq.~(\ref{eq:esd}): 
\begin{equation}
    \begin{aligned}
    	\overline{\Sigma}(R)
    	=  2\overline{\rho}\int^{\infty}_{R}[1+\xi_{gm}(r)]\frac{r dr}{\sqrt{r^2-R^2}},
    	\label{eqa:esdSR}
    \end{aligned}
\end{equation}
and
\begin{equation}
    \begin{aligned}
    	\overline{\Sigma}(\leq R)=-\frac{4\overline{\rho}}{R^2}\int^R_0{y dy}\int^{\infty}_{y}[1+\xi_{gm}(r)]\frac{r dr}{\sqrt{r^2-y^2}},
    	\label{eqa:esdR}
    \end{aligned}
\end{equation}
where $\overline{\rho}$ is the averaged background density of the universe.  
    
Since Eq.~(\ref{eq:esd_shear_t}) connects the observations with $\Delta\Sigma(R)$, it provides the method to estimate the distribution of whole underlying dark matter for foreground environments in observed region by fitting observation to the halo model.  

In the following we will consider the total mass, contributing to the $\Delta\Sigma(R)$, made of two main terms: the one-halo term and two-halo term. The first term includes all the mass contained in stars, both from the central and satellite galaxies, and the dark matter main halo. The second term is the projected two-halo term that correlates the matter in other individual halos with the main host halo. In general, the contribution from one-halo term is dominated in the scales smaller than the virial radius of host halo, and the two-halo term is forced to have an effect to the scales larger than the virial radius. According to this definition the ESD can be written as: 
\begin{equation}
	    \Delta\Sigma(R)=\Delta\Sigma_{1h}(R)+\Delta\Sigma_{2h}(R),
\end{equation}
which does not contain the contribution from the averaged background density of the universe, which does not give any contribution to the ESD, by definition.

\subsubsection{One-halo term}
The contributions from one-halo term all are given by the mass elements inside of the host halo. Specifically, $\Delta\Sigma_{1h}(R)$ is given by the three components: 1) the stellar mass density of the central galaxy, $\Delta\Sigma_*(R)$, 2) the dark matter density of the central halo, $\Delta\Sigma_{cen}(R)$, and 3) the mass density of the satellite galaxies, $\Delta\Sigma_{sat}(R)$.

The stellar component, $\Delta\Sigma_*(R)$, assumes the central galaxy as a point mass, and it could be modelled as   
\begin{equation}
    	\Delta\Sigma_*(R)=\frac{M_*}{2\pi R^2},
\end{equation}
where $M_*$ is the medians of stellar mass of galaxies from different lens samples in our specific sample. Since we measure the weak lensing signal starting from a distance that is generally a few tens of kpc from the centre, the point-mass assumption is fairly reasonable. 

For the other two components, such as the central dark halo and the overall satellite mass density, $\Delta\Sigma_{cen}(R)$ and $\Delta\Sigma_{sat}(R)$, we adopt a \cite{navarro1997} (hereafter NFW) density profile 
\begin{equation}
    	\rho(r)=\frac{\rho_0}{(r/r_s)(1+r/r_s)^2},
    	\label{eq:nfw}
\end{equation}
where $r$ is the distance from the halo centre, $r_s$ is the characteristic radius, and 
\begin{equation}
        \rho_0=\frac{\overline{\rho}\Delta_{vir}}{3I},\quad I=\frac{1}{c^3}\int^c_0 \frac{xdx}{(1+x)^2},
\end{equation}
where we have further defined mean density of $\Delta_{vir}=200$ times the critical density of the universe and a concentration parameter $c=c_{200}=r_{200}/r_s$, where $r_{200}$ is the viral radius of the halo. 

In Eqs.~(\ref{eqa:esdSR}) and (\ref{eqa:esdR}), we simply replace the $\overline{\rho}(1+\xi_{gm}(r))$ with the density distribution of the host halo $\rho(r)$ as in Eq.~(\ref{eq:nfw}). 
The projected ESD $\Delta\Sigma_{cen}$ \citep{yang2006} produced from the lensing signals $\gamma_t$ around foreground central galaxies for an NFW profile is given by
\begin{equation}
    	\Delta\Sigma_{cen}(R)=\frac{M_h}{2\pi r^2_s}I^{-1}[g(R/r_s)-f(R/r_s)],
\end{equation}
where the halo mass  $M_h=(4\pi/3)\Delta_{vir}\overline{\rho}r^3_{200}$,
\begin{equation}
    \label{eq:func1}
    f(x)=\left\{\begin{array}{ll}
    \frac{1}{x^{2}-1}\left[1-\frac{\ln \left(\frac{1+\sqrt{1-x^{2}}}{x}\right)}{\sqrt{1-x^{2}}}\right] & \text {, } x<1. \\
    \frac{1}{3} & \text {, } x=1. \\
    \frac{1}{x^{2}-1}\left[1-\frac{\operatorname{\arctan}\left(\sqrt{x^{2}-1}\right)}{\sqrt{x^{2}-1}}\right] & \text {, } x>1.
    \end{array}\right.
\end{equation}
and
\begin{equation}
    g(x)=\left\{\begin{array}{ll}
    \frac{2}{x^{2}}\left[\ln \left(\frac{x}{2}\right)+\frac{\ln \left(\frac{1+\sqrt{1-x^{2}}}{x}\right)}{\sqrt{1-x^{2}}}\right] & \text {, } x<1. \\
    2+2 \ln \left(\frac{1}{2}\right) & \text {, } x=1. \\
    \frac{2}{x^{2}}\left[\ln \left(\frac{x}{2}\right)+\frac{\operatorname{\arctan}\left(\sqrt{x^{2}-1}\right)}{\sqrt{x^{2}-1}}\right] & \text {, } x>1.
    \end{array}\right.
\end{equation}
with $x=R/r_s$.

With regard to the satellite component,
$\Delta\Sigma_{sat}(R)$, this is further composed of two contributions: first, the ESD contributed from the satellite galaxy's own host halo, $\Delta\Sigma_{s,host}(R|R_{sat})$, and second, the dark matter sub-halo, $\Delta\Sigma_{s,sub}(R)$. Hence the total satellite ESD can be written as:
\begin{equation}
    	\Delta\Sigma_{sat}(R|R_{sat})=\Delta\Sigma_{s,host}(R|R_{sat})+\Delta\Sigma_{s,sub}(R),
\end{equation}
where $R_{sat}$ is the projected off-centre distance between the satellite galaxy, which is located at the centre of its sub-halo, with the centre of its host halo.
The projected surface mass density of the host halo around a satellite galaxy at $R_{sat}$ can be given by
\begin{equation}
    	\Sigma_{s,host}(R|R_{sat})=\frac{1}{2\pi}\int^{2\pi}_0\Sigma_{\rm NFW}\sqrt{R^2_{sat}+R^2+2R_{sat}R\cos{\theta}}\ d\theta,
\end{equation}
where $\Sigma_{\rm NFW}$ is the projected density profile of the host halo.
    
According to the Eq.~(\ref{eq:esd}), we can calculate the ESD of the satellite galaxy's own host halo around satellite galaxies by
\begin{equation}
        \Delta\Sigma_{s,host}(R|R_{sat})=\Sigma_{s,host}(R\leq R_{sat})-\Sigma_{s,host}(R|R_{sat}),
\end{equation}
where the $\Sigma_{s,host}(R\leq R_{sat})$ can be derived by the integral of the $\Sigma_{s,host}(R|R_{sat})$ from 0 to $R$. And the sub-halo contribution is derived from the density profiles of stripped dark matter sub-halos that describes in \cite{Hayashi2003}.
    
In our model, we apply a simple power-law HOD for the satellite occupation function as have been studied in \cite{Mandelbaum2005b,Mandelbaum2009} assuming and NFW profile of the satellite distribution so that
\begin{equation}
    \begin{aligned}
      \Delta\Sigma_{sat}(R)=&\int_{0}^{\infty} n(M_h) \langle N_{sat} \rangle(M_h) dM_h  \\ 
      & \int dR_{sat}P(R_{sat}|M_h)\Sigma_{s,host}(R|R_{sat},M_h),\\ 
    \end{aligned}
\end{equation}
where $P(R_{sat}|M_h)$ is simply the $f(x)$ in the Eq.~(\ref{eq:func1}). The $ \langle N_{sat} \rangle(M_h)$ is the occupation function of satellite galaxies given a halo mass $M_h$. The $n(M_h)$ is the halo mass function based on \cite{tinker2005}.

Finally, the one-halo term is composed of the stellar mass contribution with the dark halo contributions, which weighted by satellite fraction, from the central and satellite galaxies:
\begin{equation}
    	\Delta\Sigma_{1h}=\Delta\Sigma_*+(1-f_{sat})\Delta\Sigma_{cen}+f_{sat}\Delta\Sigma_{sat}.
\end{equation}
The halo mass $M_h$ is mostly provided by the total mass of one-halo term that embraces the baryons and the NFW virial mass $M_{200}$, which represents the mean density 200 times the critical density within the radius $r_{200}$.

\subsubsection{Two-halo term}
The two-halo term arises from the matter of satellite galaxies in parted halos that are correlated with the large scale distribution of dark matter in the host halos \citep{yang2006}. As the scale increases, the two-halo term is supposed to gradually dominate the ESD signals. In order to obtain the ESD from the contribution of two-halo term, we apply $pyCamb$ \citep{Lewis2013} to calculate the power spectrum at the median redshift of each sample. Then the matter-matter correlation function $\xi_{mm}$ can be converted by the power spectrum. Next, we use $\xi_{mm}$ to calculate the halo-matter correlated function $\xi_{hm}$ by the scale-dependent bias model \citep{tinker2005},
\begin{equation}
       \xi_{hm}=b_h(M_h)\eta \xi_{mm},
\end{equation}
where 
\begin{equation}
        \eta(r)=\frac{(1+1.17\xi_{mm}(r))^{1.49}}{(1+0.69\xi_{mm}(r))^{2.09}},
\end{equation}
and $b_h(M_h)$ is the halo bias  \citep{seljak2004} as the function of the halo mass. Then the two-halo term $\Delta\Sigma_{2h}$ can be estimated by using the Eq.~(\ref{eq:esd}).

\subsection{Fitting Process}
\label{sec:fitting}
Given the halo model defined above, the measured ESDs is used to constrain the halo properties of the corresponding lens sample. Specifically we want to constrain the three free parameters of the model, such as the virial halo mass, $M_h$, and the concentration, $c$, of the one-halo term, and the satellite fraction, $f_{sat}$.

To best fit the observed EDS, we use the $\verb|emcee|$ \citep{emcee} python pipeline, which makes use of a standard Markov Chain Montecarlo (MCMC) procedure \citep{wentao2022} to explore the likelihood function in the multidimensional parameter space. The maximum likelihood function is a Gaussian where the covariance matrix is estimated by bootstrap sampling. In our analysis, we have adopted a flat prior distribution with the host halo mass $log({M_h}/M_{\odot})$ in the range [9.5, 14.0], the concentration $c$ in the range [0.1, 20.0], and the satellite fraction, $f_{sat}$ in the range [0.0, 1.0]. We set a rather broad range for the parameters space for reducing the prior effects as far as possible.


\section{Results}
\subsection{Systematics Tests}
\label{sec:systematics}
The assessment of systematics is a crucial part of the galaxy-galaxy lensing analyses, as systematics impact the reliability of the results. 

\begin{figure}
    \centering
    	\resizebox{\hsize}{!}{\includegraphics{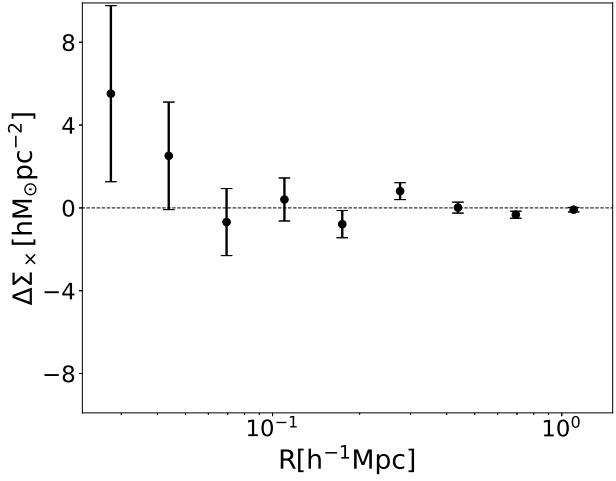}}
    	\caption{B-mode systematics test in the galaxy-galaxy lensing measurements. The black points with error-bars represent the 'B-mode signals' $\Delta\Sigma_\times$, the cross component of lensing signals from the background sources, measured around the FLS of VOICE.  }
    	\label{fig:Bmode_full}
\end{figure}

The first test is related to the B-mode signal. This represents the cross components of the galaxy-galaxy lensing signals, $\gamma_\times$, along a direction tilted by $45^\circ$ with respect to the tangential component, $\gamma_t$. The $\gamma_t$ produces itself a ESD cross component, $\Delta\Sigma_\times$, tilted with respect to the tangential components, the $\Delta\Sigma$. 
By definition, the B-mode signal is zero for an unbiased shear signal. Thus, any deviation from zero can indicate the presence of systematics in the ESD, which is diluted by the presence of a off-axis shear. Fig.~\ref{fig:Bmode_full} shows the B-mode signals $\Delta\Sigma_\times$ for the VOICE FLS. 
This is generally consistent with zero for all scales as expected for lensing though the most inner $\Delta\Sigma_\times$ point is somehow deviated from zero.
The B-mode tests for the Red/Blue Lens sub-samples is discussed in details in Appendix.~\ref{app:A}. These also show basically no systematics, althought the error bars become, in some cases, rather large.  
Since all $\Delta\Sigma_\times$ are almost all  consistent with zero, we conclude that the systematics, if any, are reasonably confined within the statistical errors. 

\begin{figure}
    \centering
    	\resizebox{\hsize}{!}{\includegraphics{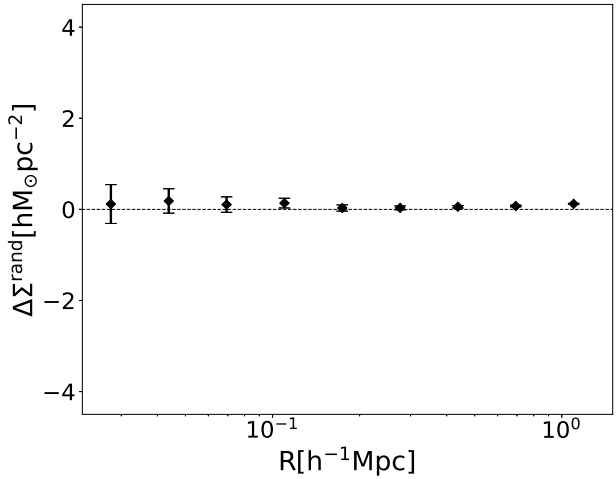}}
    	\caption{Random-points systematic test in the galaxy-galaxy lensing measurements. The black points with error-bars represent the $\Delta\Sigma^{\rm rand}$, the tangential component of lensing signals from the same sources sample, measured around points of random lens sample correspond to FLS. }
    	\label{fig:rand_test_full}
\end{figure}

We generate lens samples of random points as 100 times of the numbers correspond to the FLS, and Red/Blue Lens sub-samples, respectively. Fig.~\ref{fig:rand_test_full} shows the ESD signals $\Delta\Sigma^{\rm rand}$ of  background sources, measured around the random points from FLS. This is, again, fairly consistent with zero overall, with marginal evidence of a positive signal in the first bin. The results for the Red/Blue Lens sub-samples are discussed in details in Appendix.~\ref{app:A}. They show a similar pattern, with random signal generally consistent with zero.

\begin{figure}
    \centering
	    \resizebox{\hsize}{!}{\includegraphics{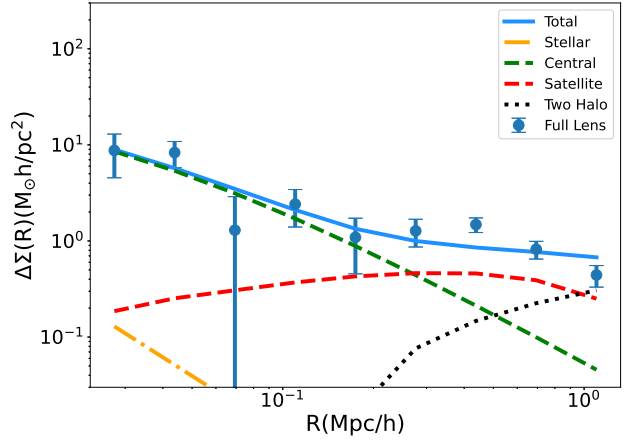}}
	    \caption{ESD signals ($\Delta\Sigma$) around FLS galaxies (blue points with bars), and the best-fitting curve (blue solid line) is comprised of the contribution from different components, which contain stellar term (orange dash-dotted line), central term (green dashed line), satellite term (red dashed line), and two halo term (black dotted line). }
    	\label{fig:sigma_fitting_lines_full}
\end{figure}

\begin{figure}
    \centering
    	\includegraphics[width=8.5cm]{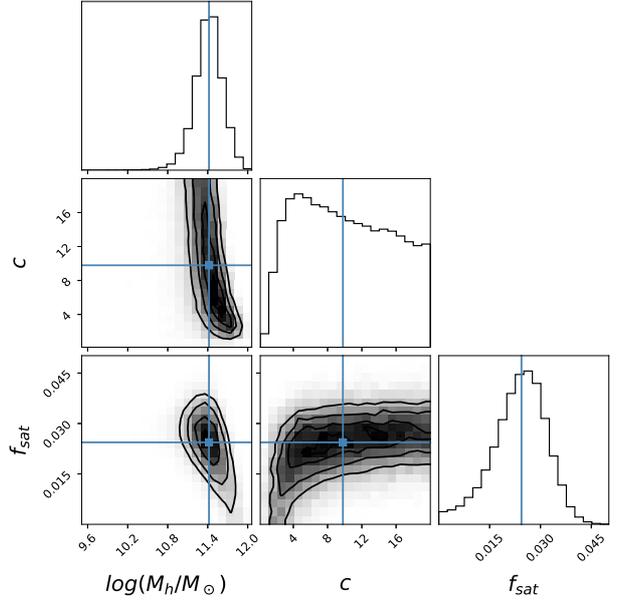}
    	\caption{Marginalised posterior distributions of three parameters obtained by using MCMC method for halo model to fit our ESD measurements around the galaxies of FLS. The three contour levels correspond to 16$\%$, 50$\%$, 84$\%$ confidence levels, respectively. The blue points and lines are the medians. }
    	\label{fig:model_constrains_full}
\end{figure}

    
A final note of caution is needed about the innermost bin at $\sim27$kpc h$^{-1}$, corresponding to $\sim9''$ in angular scale. 
In both tests above, we have stressed a marginal systematic deviation of the $\Delta\Sigma_\times$ and $\Delta\Sigma^{\rm rand}$ from zero. There are two possible explanation that can mitigate the impact of this source of systematic.
On the one hand, the lens sample dominated by faint galaxies with low stellar mass is different with \citealt{Sifon2018} which finds an additive bias as a bright lens influences the shapes of background sources in small scale. So, the effect is negligible for these faint galaxies of our lens samples in this small scale.
On the other hand, the $\Delta\Sigma_\times$ and $\Delta\Sigma^{\rm rand}$ represent the systematics which should tend to zero, but it is reasonable that both are deviated from zero if the counts of lens-source pairs decrease. The fact is that the area of innermost bin is smaller than outer bins. Therefore, there are less counts of sources around lens in the smaller scale that cause the bigger deviation from zero for the innermost ESD measurements. 
However, the $\Delta\Sigma_\times$ and $\Delta\Sigma^{\rm rand}$ for all radial bins are within 2$\sigma$ represent there are acceptable systematics so that we would keep the innermost ESD measurements.

\subsection{Halo property Constraints}
\label{sec:halo_constraints}
In this section we present the halo model constraints of the VOICE lens samples using the MCMC procedure introduced in Sect.~\ref{sec:fitting} to fit the ESD signals produced by the source sample.

\begin{table*}\footnotesize
\renewcommand\arraystretch{1.6}
	\centering
	\caption{Posterior constraints derived from the best fitting to the  $\Delta\Sigma$ measured around the Blue Lens-1/2, Red Lens-1/2, and Full Lens by our halo model with responding reduced $\chi^2$ and p-value. It presents the median of logarithmic stellar mass $\log{(M_*/M_\odot)}$, and the median parameter constraints: $\log{(M_h/M_\odot)}$, $c$, $f_{sat}$ with statistical errors. The reduced $\chi^2$ and $p$-value in brackets indicate the measurements fit the model without the outermost data point. }
	\label{Tab:02}
	\begin{tabular}{lcccccccr}
		\hline\hline
		Lens Sample & $\log{(M_*/M_\odot)}$ & $\log{(M_{h}/M_\odot)}$ & $c$ & $f_{sat}$ &$\chi^2_{d.o.f=5\ (d.o.f=4)}$&$p${-}value\\
		\hline
		{$Blue\quad Lens-1$} & 8.31 & 11.24$^{+0.20}_{-0.31}$ & 10.6$^{+6.1}_{-5.6}$ & 0.004$^{+0.005}_{-0.003}$ & 3.539 (1.644) & 0.003 (0.160)\\

		{$Blue\quad Lens-2$} & $9.79$ & 11.61$^{+0.44}_{-0.87}$ & 9.3$^{+7.1}_{-6.2}$ & 0.010$^{+0.003}_{-0.003}$ & 0.203 & 0.961\\
		\hline
		{$Red\quad Lens-1$} & $10.01$ & 11.84$^{+0.34}_{-0.69}$ & 10.9$^{+6.2}_{-6.6}$ & 0.128$^{+0.034}_{-0.037}$ & 1.253 &0.281\\

		{$Red\quad Lens-2$} & $10.73$ & 12.71$^{+0.22}_{-0.30}$ & 9.3$^{+6.4}_{-4.7}$ & 0.236$^{+0.087}_{-0.105}$ & 0.926 & 0.463\\
		\hline
		{$Full\quad Lens$} &  $8.49$ & 11.42$^{+0.19}_{-0.20}$ & 9.8$^{+6.6}_{-5.6}$ & 0.024$^{+0.007}_{-0.008}$ & 1.995 & 0.076 \\		
		\hline
	\end{tabular}

\end{table*} 


\begin{figure*}
	\centering
	\includegraphics[width=17cm]{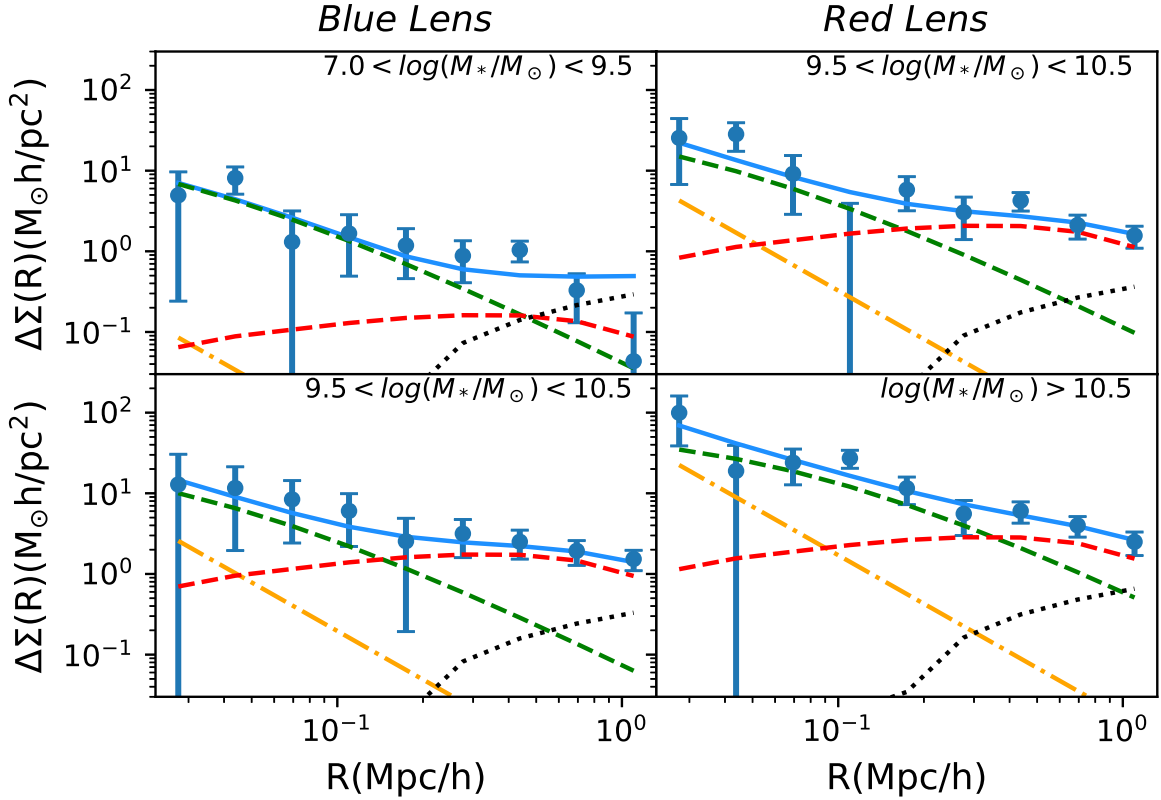}
	\caption{Model fitting curves from MCMC procedure for the $\Delta\Sigma$ signals measured around Blue Lens-1 (top left panel), Blue Lens-2 (bottom left panel), Red Lens-1 (top right panel), and Red Lens-2 Lens (bottom right panel). The best-fitting curve (blue solid line) is comprised of the contribution from different components that are stellar term (orange dash-dotted line), central term (green dashed line), satellite term (red dashed line), and two-halo term (black dotted line), respectively. }
	\label{fig:subs_sigma_model}
\end{figure*}

\begin{figure}
    \centering
    	\resizebox{\hsize}{!}{\includegraphics{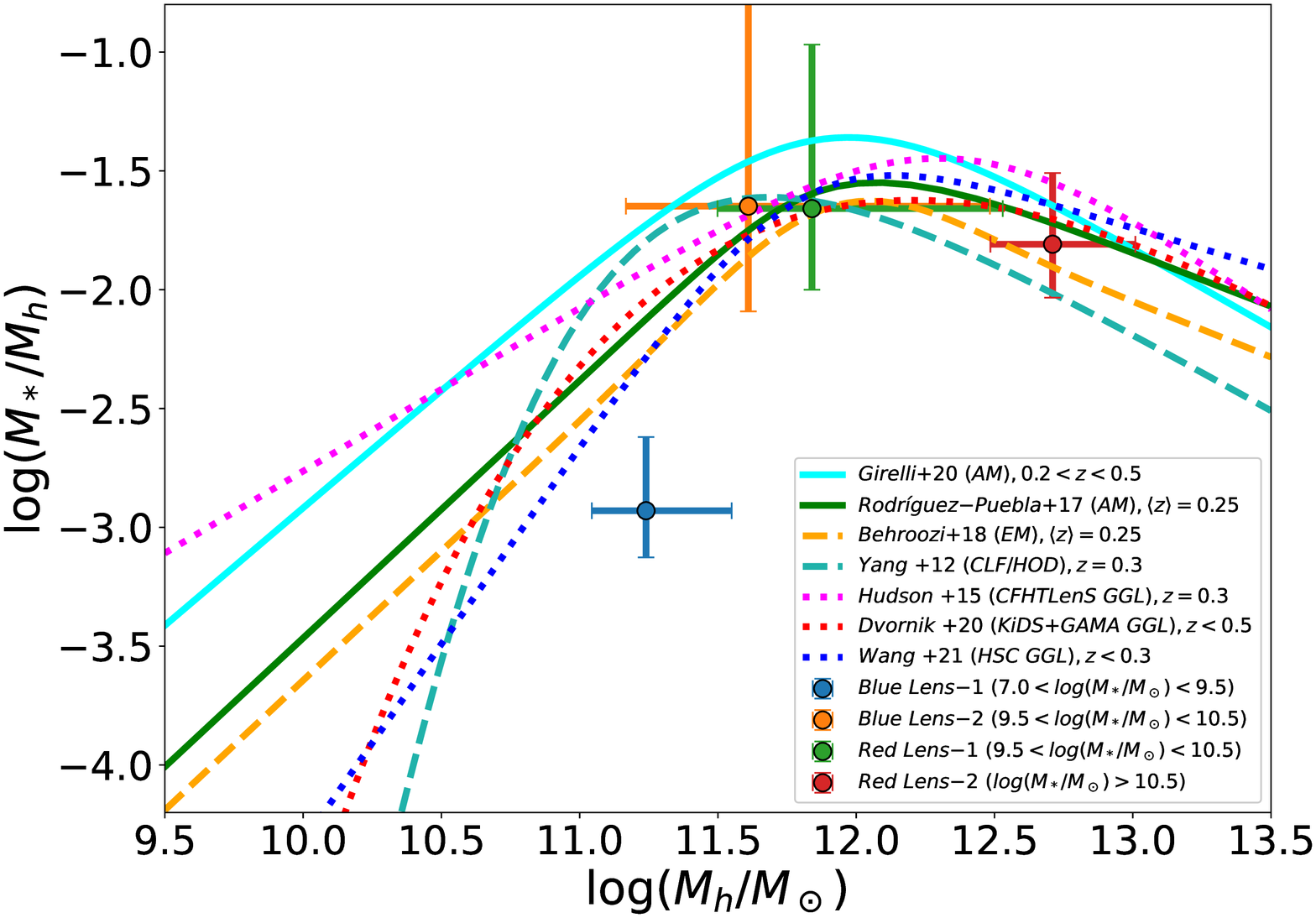}}
    	\caption{Relation of  the fraction of stellar and halo mass versus halo mass. The red, green, orange, blue circular points with cross error bars are the SHMR results of Red/Blue Lens$-$1/2, respectively. For comparison, here displayed are the SHMR results from different models: AM (the cyan and green solid lines), EM (the orange dashed line),  and CLF/HOD(the sea-blue dashed line). Also displayed are galaxy-galaxy lensing results from the surveys of CFHTLenS (the magenta dotted line), KiDS+GAMA (the red dotted line), and HSC (the blue dotted line). }
    	\label{fig:SHMR_subsamples}
    \end{figure}

The results of $\Delta\Sigma$ of the FLS are shown as blue points with error bars in Fig.~\ref{fig:sigma_fitting_lines_full}. The error bars of $\Delta\Sigma$ are calculated from the standard errors based on the bootstrap via re-sampling $10^4$ times. 
In the same figure, the blue solid curve is the best fit line to the ESDs of the FLS, given by the total model, as the sum of all contribution of the different mass components, defined by the free parameters.
In details: 1) the orange dash-dotted line represents the contribution from stellar mass of foreground galaxies, which is defined by the mean stellar mass derived from the stellar population analysis; 2) the green dashed line is the NFW model defined by the best fit parameters $c$ and $M_h$; 
3) the red dashed line represents the satellite galaxies, defined by the other free parameter $f_{\rm sat}$;
4) the black dotted line describes the contribution from two-halo term. These contributions all are described in Sect.~\ref{sec:halo_model}.

From Fig. \ref{fig:sigma_fitting_lines_full}, it is clear that the satellite component and the two-halo term dominate the large scale, while the stellar mass and mostly the dark halo dominate on small scales. Overall, the total fit is reasonably good with a reduced $\chi^2\approx$ 1.995 (d.o.f=5, see Table~\ref{Tab:02}).

The marginalised posterior distributions of the three parameters obtained for the FLS sample is shown in Fig.~\ref{fig:model_constrains_full}. The three contours correspond, from the innermost to the outermost one, to $16\%$, $50\%$, $84\%$ confidence levels. For the FLS, the median of host halo mass is 
$10^{11.42}\ M_\odot$, which, compared to the stellar mass $M_*$, implies a $M_h/M_*=10^{2.93}$.
It is evident, from both $M_h$ and $M_*$, that the sample is dominated by low mass systems, as also discussed in Sect.~\ref{sec:lens_sample}.

To explore the stellar to dark matter relation, we proceed to best fit also the other samples split by mass and colours, as defined in Sect.~\ref{sec:lens_sample}. In Fig.~\ref{fig:subs_sigma_model} we show the best fit models with contributions of different components for the $\Delta\Sigma$ from Red/Blue Lens sub-samples in the different mass bins adopted. We found the corrected $\Delta\Sigma$ of Blue Lens-1 dropped at the large scale due to the subtraction. The value of the blue low mass bin signal ($\Delta\Sigma=$0.18) is too small for the subtraction which means it is sensitive to $\Delta\Sigma^{rand}$ at the large scale for low mass lenses. So we measure the reduced chi-square $\chi^2_{d.o.f}$ with or without the last data point (d.o.f = 5 or 4) after subtracting the outermost data point in Table~\ref{Tab:02}. And the outermost data point does not change the halo mass significantly, and we think the model fitting is generally consistent with data points cause the $\chi^2_{d.o.f=4}$ is 1.644 and p-value is 0.160.

Here we can appreciate the variance of the $\Delta\Sigma$ amplitude as a function of the sample mass. In particular the central peak of the most massive (Red) sample (bottom-right panel) is one order of magnitude larger than the one of the least (blue) massive sample. Looking at the typical systematics from the cross and random samples (see Appendix.~\ref{app:A}), it is evident that these are negligible with respect to the signal of the massive samples ($\log M_*/M_\odot>9.5$), while they migh affect the low mass sample ($\log M_*/M_\odot<9.5$). Another evident feature is that the stellar component seems to be more centrally concentrated for the massive systems, with respect to the less massive ones, while the satellite fraction decreases with the stellar mass (see also Table~\ref{Tab:02}). 

Overall, the total model allows us to fit rather well the ESD of all sample with reduced $\chi^2$ smaller than the FLS (see Table~\ref{Tab:02}). The halo masses $M_h$ of the four samples have positive relation with the stellar masses $M_*$ that is physically reasonable.
On the other hand, the concentrations $c$ does not show a clear (anti) correlation with the virial mass, as predicted from simulations (e.g. \citealt{neto2007}). 
    
To conclude this section, we compare the Stellar-to-Halo Mass Relations (SHMR) of our results with literature. Fig.~\ref{fig:SHMR_subsamples} shows the comparisons between the SHMR results from Red/Blue Lens sub-samples in the similar redshift ranges with seven different curves which are the results from three models and galaxy-galaxy lensing (GGL) analyses of three survey: abundance matching (AM; \citealt{Girelli2020}, \citealt{Aldo2017}), empirical modelling (EM; \citealt{Behroozi2019}), and the conditional luminosity function or halo occupation distribution (CLF/HOD; \citealt{yang2012}); CFHTLenS \citep{hudson2015}, KiDS+GAMA \citep{Dvornik2020} and HSC \citep{Wang2021}. 
As we can see in Fig.~\ref{fig:SHMR_subsamples}, the SHMRs of Blue Lens$\verb|-|$2 and Red Lens$\verb|-|1\&2$ have good agreements with the results from other studies, but it is situated below these curves for Blue Lens$\verb|-|$1 that means the stellar mass is lower than these predictions under the certain low halo mass.

  
\section{Discussion and Conclusion}

We measure the galaxy-galaxy lensing signals around galaxies selected from VOICE photometric catalogue by stacking the background galaxy shape behind them. The shape catalogue is based on the full VOICE photometric catalogue but with selection criteria designed for weak lensing analysis as described in F18. In this section, we discuss our major results and draw some conclusions. 

The 4.9 deg$^2$ multi-band VOICE deep imaging survey overlaps with Chandra Deep Field-South (CDFS). The depth is down to 26.1 (5$\sigma$ limiting magnitude) in the $r$ band. We select the full lens sample (FLS) between redshift 0.1$<z_{l}(\verb|BPZ|)<0.35$ and further split it into four sub-samples based on stellar mass and colour. During the stacking process, we select the background galaxies to be at a higher redshift than the lens sample, such as $z_s>z_l+0.2$, to avoid contamination from the unlensed galaxies. A boost factor has been applied to each measurement as a correction from the residual contamination induced by the effect of lens-sources physical correlation.

A series of tests have been done to assess the systematics in the measurements, including the B-mode test and random samples test. Both null tests are consistent with zero for all the samples except the ESDs of innermost radial bins. 
It is acceptable that these innermost ESDs are within $1\sim2\sigma$ for FLS and Blue Lens samples and within $1\sigma$ for Red Lens samples.
We also cross test the reliability using the DES$\verb|-|$Y1 public shape catalogue in the VOICE region, we find the results are consistent with each other, regardless of the different shape measurement methods ( $\verb|LensFit|$ for VOICE and $\verb|METACALIBRATION|$ for DES). Except that the DES measurements are noisier than VOICE measurement due to the shallower survey depth of DES (See. Appendix~\ref{app:B}). 

Due to the fact that $\Sigma_{\rm crit}$ depends on the photo-z of sources around each lens, the redshift uncertainties would influence the ESD measurement. In order to test this effect, we select sources around each lens where the accumulated probability of the photo-z of each source satisfies the requirement \citep{Medezinski2018}: $P(z_s>z_l+0.2)>0.98$, and then we correct $\Sigma_{\rm crit}$ by applying the $P(z_s)$ to marginalise over photo-z errors according to \cite{wentao2022}. 
However, this selection leads to larger statistical errors in the ESD measurements, since it significantly reduces the number of sources around each lens. We have checked and got consistent ESD measurements between using (See. Appendix~\ref{app:C}) and not using (See. Sect.~\ref{sec:halo_constraints}) this method for our samples, but the bootstrap error bars of the former ESD measurements are larger than that of the latter. We decide not to apply the full p(z) in the ESD measurement for our samples. On the other hand, F18 has presented the photo-z ($\verb|BPZ|$) accuracy in detail, and we think the photo-z uncertainties of sources are enough.

We then fit the ESDs with a three parameter model, such as halo mass, concentration of the halo and satellite fraction. For the FLS, we estimate the halo mass to be $M_h=10^{11.42^{+0.19}_{-0.20}}M_\odot$, the concentration parameter $c=9.8^{+6.6}_{-5.6}$, and the satellite fraction $f_{sat}=0.004^{+0.005}_{-0.003}$. For the sub lens samples, we check the stellar mass to halo mass relation (SHMR) and compare our results to various existing SHMR models (Fig. \ref{fig:SHMR_subsamples}). 

We find that the blue low-mass lens sample Blue Lens-1 (median $M_*=10^{8.31} M_\odot$) have significantly larger halo mass than theoretical prediction while the others are consistent with theory. This result actually agrees with what have been found in \cite{hudson2015}, who found that the mass to light ratio of the faint blue dwarfs deviates towards higher value  than the abundance matching prediction. \cite{boylan-kolchin2012MNRAS} also showed similar results for low-mass galaxies from dynamical analysis. Rotation curve analysis by \cite{ferrero2012MNRAS} again obtain similar results for dwarf galaxies. Interestingly, this seems the opposite for massive star forming galaxies where the halo mass is much lower than the theoretical predictions by \cite{zhang2021arXiv211204777Z} indicating a very high gas to stellar conversion rate (up to 67\%) at stellar mass around $10^{10.75} M_{\odot}$.

Another interesting finding of our work is that the blue dwarf galaxies, such as Blue Lens-1 sample, occupies  $\sim$75$\%$ of the full sample, implying the VOICE-CDFS region dominated by low-mass blue dwarf galaxies in redshift range of our FLS. This result agrees with the work from \cite{Phleps2007} who estimated the overdensities of three COMBO-17 fields,  and found that the CDFS region density is 2 times lower than the other two regions which agrees with local 2dF observations. We then suggest that any cosmological constraints using the data in this region may suffer from sever cosmic variance in this particular redshift range. 

As the first paper of galaxy-galaxy lensing study from VOICE deep imaging data, we test the robustness of our measurement and obtain the SHMR for various of lens galaxy sample. The deep imaging data enables galaxy-galaxy lensing analysis for even higher redshift objects. In future, combining X-ray data from CDFS we will further explore the the halo properties of those X-ray selected AGNs using VOICE shape catalogue (Li et al in prep).

\begin{acknowledgements}
      The corresponding authors are Liping Fu, Wentao Luo, and Nicola R. Napolitano. RL acknowledges the supports from Sun Yat-sen University and Shanghai Normal University for the numerical computations are carried out through Kunlun cluster of SYSU and the computer cluster of SHNU. LF acknowledges the supports from NSFC grants 11933002, and the Dawn Program 19SG41 \& the Innovation Program 2019-01-07-00-02-E00032 of SMEC. WL is supported in part by the National Key R\&D Program of China (2021YFC2203100), the NSFC (No.11833005, 12192224). We acknowledge the science resarch grants from the China Manned Space Project with No. CMS-CSST-2021-A01. N.R.N acknowledge financial support from the “One hundred top talent program of Sun Yat-sen University” grant N. 71000-18841229. MV acknowledges financial support from the Inter-University Institute for Data Intensive Astronomy (IDIA), a partnership of the University of Cape Town, the University of Pretoria, the University of the Western Cape and the South African Radio Astronomy Observatory, and from the South African Department of Science and Innovation's National Research Foundation under the ISARP RADIOSKY2020 Joint
      Research Scheme (DSI-NRF Grant Number 113121) and the CSUR HIPPO Project (DSI-NRF Grant Number 121291).
       
\end{acknowledgements}

%
%



\appendix

\section{Systematics for Lens Sub-samples}
\label{app:A}
\begin{figure*}
	\centering
	\includegraphics[width=16.8cm]{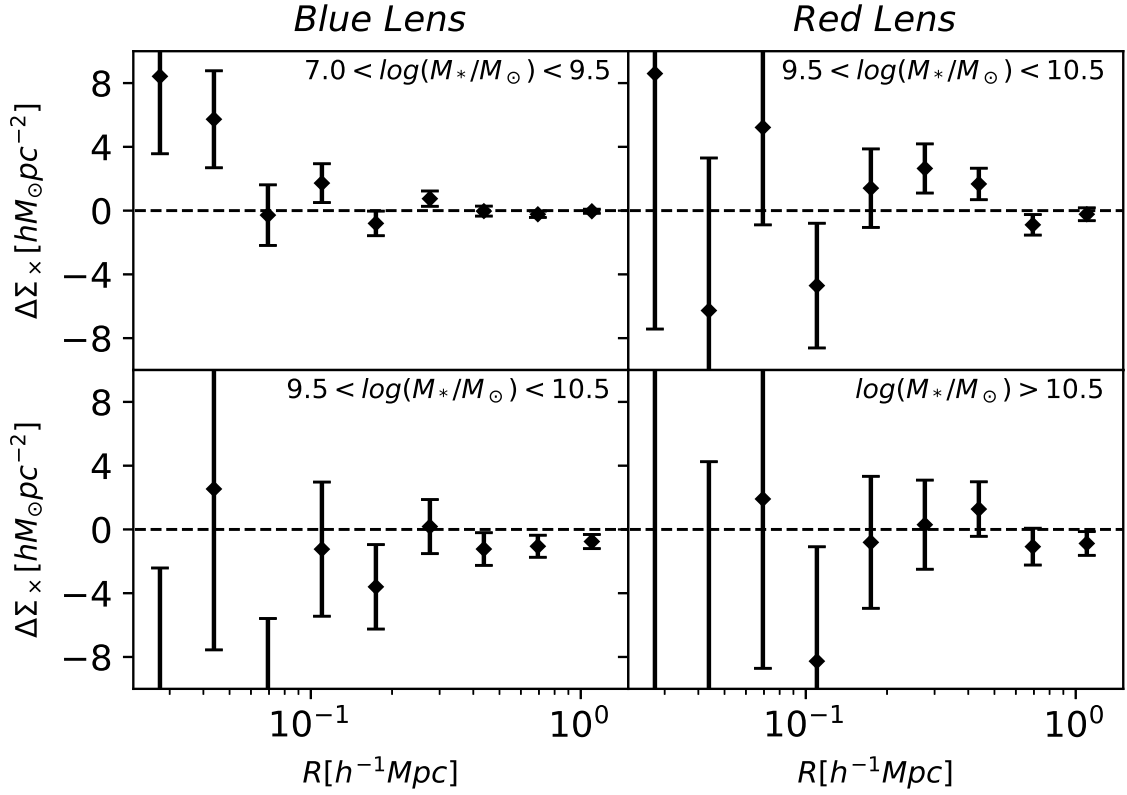}
	\caption{Top/Bottom left panels show the cross-components of $\Delta\Sigma$ measured around the galaxies of Blue Lens-1/2. And the $\Delta\Sigma_\times$ in the top/bottom right panels are measured around the galaxies of Red Lens-1/2.}
	\label{fig:a1}
\end{figure*}

\begin{figure*}
	\centering
	\includegraphics[width=16.8cm]{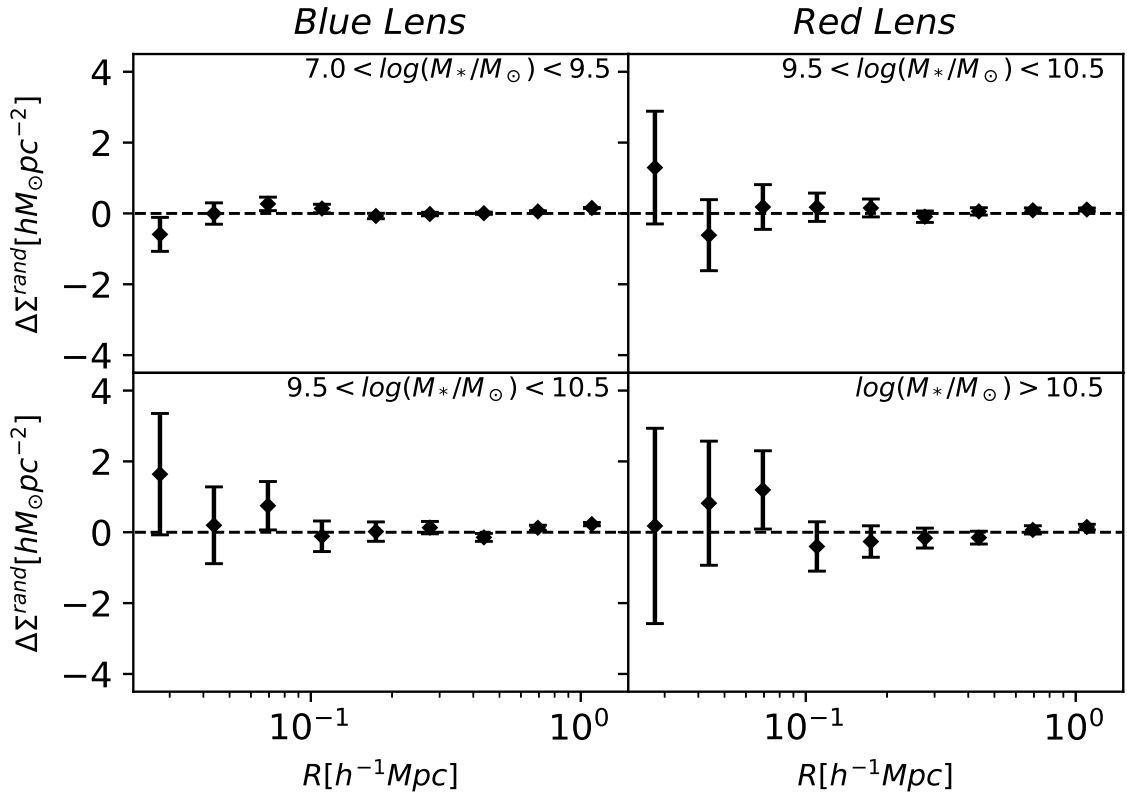}
	\caption{Top$\&$bottom left, top$\&$bottom right panels show the $\Delta\Sigma^{\rm rand}$ measured around random samples that are 100 times numbers of Blue Lens-1$\&$2 and Red-1$\&$2 samples, respectively. }
	\label{fig:a2}
\end{figure*}

In the Sect.~\ref{sec:lens_sample}, we separate the FLS into Blue Lens$\verb|-|$1/2 and Red Lens$\verb|-|$1/2 according to corresponding classification. Here we show some results of systematics tests and ESD measurements around the foreground galaxies Red/Blue Lens sub-samples based on the same background sources sample of VOICE. 
We have measured the systematics of cross components $\Delta\Sigma_{\times}$ for the four lens samples in Fig.~\ref{fig:a1} and of $\Delta\Sigma^{\rm rand}$ around the lens galaxies of the four random points samples in Fig.~\ref{fig:a2}, respectively. It is obviously that they have big statistical errors in small scale due to the galaxy number is not much. But the results of cross components and the ESDs of random points can be thought as accepted systematics due to they are consistent with zero when the scale increases.

\section{Measurements with DES-Y1 Data}
\label{app:B}
	
\begin{figure*}
	\centering
	\subfigure
	{
	    \begin{minipage}[t]{0.45\textwidth}
	    \centering
	    \resizebox{\hsize}{!}{\includegraphics{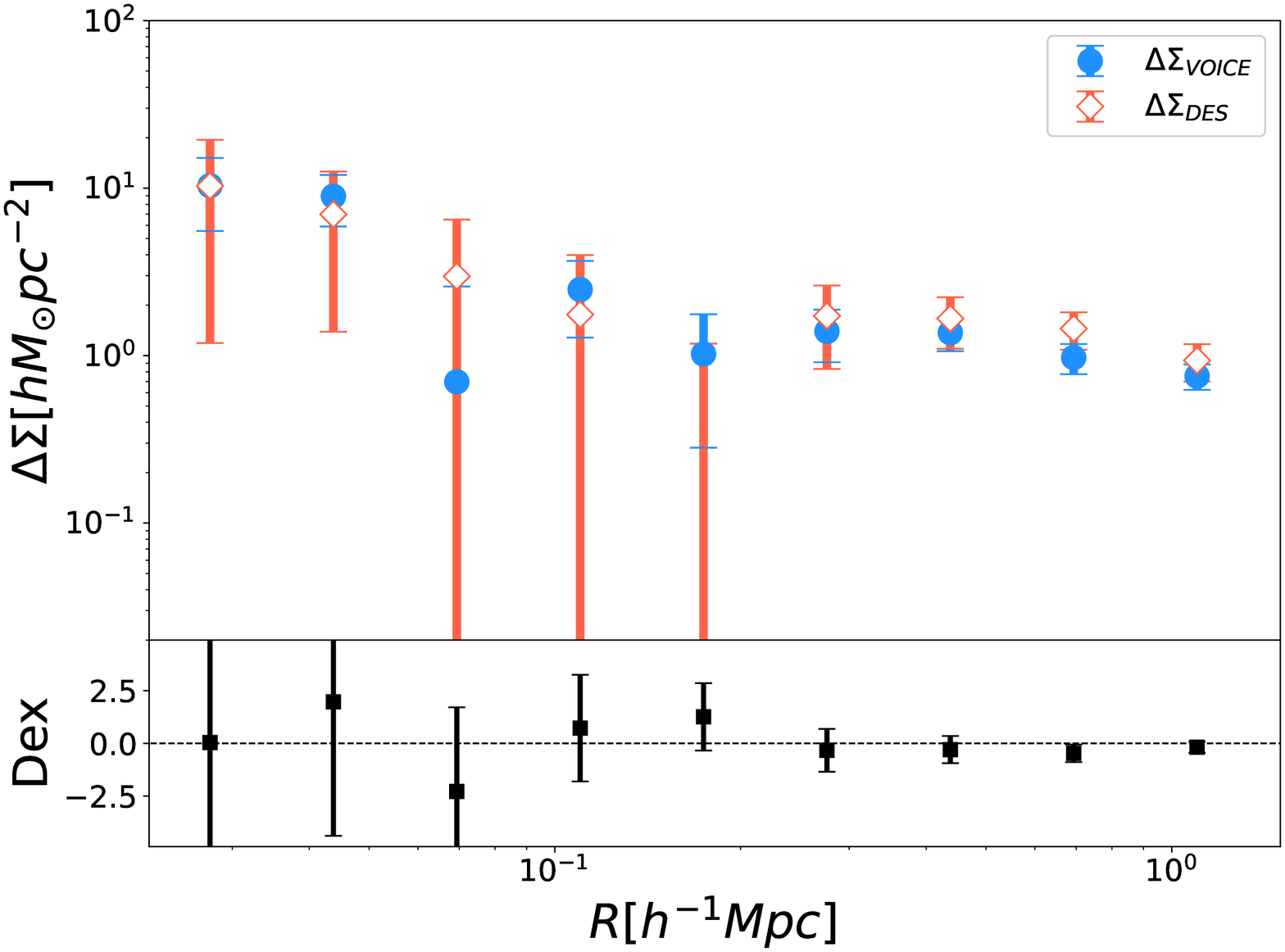}}
	    \end{minipage}
	}
	\hspace{0.2in}
	\subfigure
	{
	    \begin{minipage}[t]{0.45\textwidth}
	    \centering
	    \resizebox{\hsize}{!}{\includegraphics[width=8cm]{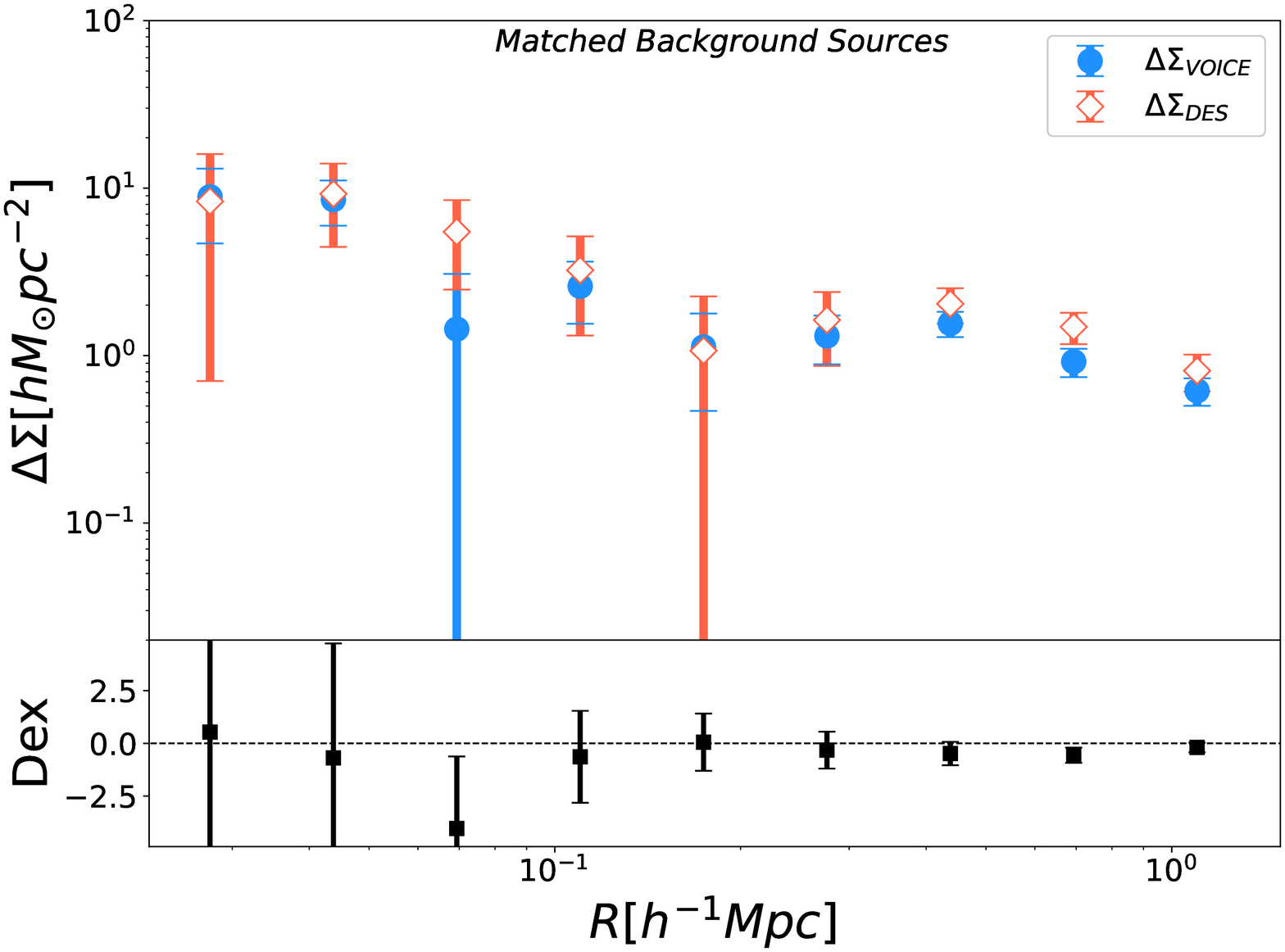}}
	    \end{minipage}
	}
	\caption{Left panel plots the ESDs around FLS with the background sources of VOICE and DES-V, and the right panel plots the ESDs around FLS with Matched Background Sources (MBS) for two surveys. The blue points and orange empty diamonds with error bars represent the results are based on the shape measurements in VOICE and DES surveys, respectively. The black points with error bars (the dex) that represent the differences that the measurements from the VOICE relative to the DES.}
	\label{fig:b1}
\end{figure*}
     
For verifying the reliability of halo properties in the VOICE$\verb|-|$ CDFS region, we compare VOICE with the results based on Dark Energy Survey (DES) Year-one annual Data\footnote{https://des.ncsa.illinois.edu} (Y1A1 or Y1) due to there are almost same but not complete coverage ($\sim93\%$) with VOICE$\verb|-|$CDFS in DES$\verb|-|$Y1. The Dark Energy Survey (DES; \citealt{DEScollaboration2016}) is a large imaging survey that uses $3\ deg^2$ Dark Energy Camera (DECam; \citealt{Flaugher2015}), a Megapixel camera installed at prime focus on the Blanco 4-m telescope at the Cerro Tololo Inter-American Observatory in northern Chile \citep{Morganson2018}. The DES survey plan to covers $\sim$1800 deg$^2$ wide-area with exposures in $grizY$ bands \citep{Drlica2018} that are less than 9 bands in VOICE survey project and provide less exposure time than VOICE survey for the corresponding band. The DES$\verb|-|$Y1 shear measurements are based on the $i$ band images with a median seeing of $\sim$0.99 arcsec \citep{Morganson2018} which is little worse than the selected exposure seeing which is $\leq$0.9 arcsec in VOICE $r$ band images. Although the depth of DES is shallower than VOICE survey, the comparison of galaxy-galaxy lensing results between VOICE and DES is a meaningful cross check to the results of halo properties in VOICE$\verb|-|$CDFS region.

\subsection{DES-Y1 Shear catalogue} 
\label{sec:B1}
Here we would use the shear catalogue from DES projects to measure ESD signal around FLS galaxies for verifying the reliability of ESD measurements from VOICE shear catalogue. \cite{Zuntz2018} introduces two independent catalogues of galaxy shape measurements from DES$\verb|-|$Y1 Data and one of them is called the $\verb|METACALIBRATION|$ \citep{Huff2017,Sheldon2017} shear catalogue\footnote{https://des.ncsa.illinois.edu/releases/y1a1/key-catalogues/key-shape} which covers 1500 deg$^2$ of the Southern sky and contains 34.8 million objects. Based on the $\verb|METACALIBRATION|$ shear catalogue, we make a sample from a cut of the area that is the almost same ranges of $RA$ and $DEC$ as the region of VOICE$\verb|-|$CDFS though the coverage of DES is not very completely overlapped. Then we can get the shear catalogue from DES$\verb|-|$Y1 Data in VOICE region (hereafter DES$\verb|-|$V) which contains 244016 galaxies in $0.3<z_s($\verb|BPZ|$)<1.5$ that is the same redshift range as the background sources sample of VOICE. The 10$\sigma$ limiting magnitude in $i$ band for DES is $\sim22.5$, and the 5$\sigma$ limiting magnitude in $r$ band for VOICE is $\sim26.1$. The significant reason why there are less galaxies in the DES-V shear catalogue is that VOICE survey can observe more faint objects than DES in the deeper space. According to \citealt{Lee2022}, we estimate averaged shear measurements by
\begin{equation}
	    \overline{\gamma}_t(\theta)=\frac{1}{\overline{\boldsymbol{R}}}\frac{\Sigma_j w'_j\gamma_{t,j}}{\Sigma_j w'_j},
\end{equation}
where $\gamma_{t,j}$ and $w'_j$ is the tangential shear measurements and the weight of source from the $\verb|METACALIBRATION|$ shear catalogue, and $\overline{\boldsymbol{R}}$ is the mean response averaged over the sources which defined as the sum of averaged measured shear response $\boldsymbol{R}_\gamma$ and shear selection bias correction matrix $\boldsymbol{R}_S$ for $\verb|METACALIBRATION|$:
\begin{equation}
        \overline{\boldsymbol{R}}=\overline{\boldsymbol{R_\gamma}}+\overline{\boldsymbol{R_S}}.
\end{equation}
Then the ESD measurements can be derived by Eqs.~(\ref{eq:esd_shear_t})-(\ref{eq:weight}). Furthermore, we make another shear catalogue by position matching for the galaxies from DES and VOICE shear catalogues that contains 148285 galaxies in $0.3<z_s($\verb|BPZ|$)<1.5$ as the Matched Background Sources (hereafter MBS).

\subsection{Comparisons of Measurements} 
\label{sec:B2}
For comparing the differences of galaxy-galaxy lensing measurements between the background galaxies of VOICE and DES for the same lens sample which we decide to use FLS, Here we do two comparisons of VOICE versus DES for checking the reliability of the ESD ($\Delta\Sigma$) measurements from VOICE and the differences between two pipelines of shape measurements in Fig.~\ref{fig:b1}. 
    
As a sanity check for the ESD measurements in the VOICE study, we measured the $\Delta\Sigma$ around FLS galaxies with the background sources from DES$\verb|-|$V shear catalogue under the selection of lens-source pair: $\Delta z_p>0.2$. Then we compare the results of $\Delta\Sigma$ in DES survey with VOICE survey in the left panel of Fig.~\ref{fig:b1} that shows the dex of difference of ESD measurements from two surveys in the almost same area is consistent with zero.  
    
We also need to consider whether there are big difference between the ESD measurements from the two kind of shape measurement pipelines: $\verb|Lensfit|$ in VOICE and $\verb|METACALIBRATION|$ in DES. Then we measured the ESDs around FLS galaxies with the MBS galaxies for the different pipelines of two surveys in the right panel of Fig.~\ref{fig:b1} that show the difference is almost consistent with zero. Since the differences of $\Delta\Sigma$ between two surveys are within 2$\sigma$, the very little discrepancy is acceptable for both ESDs . 
    
The results from DES survey to the VOICE survey are good cross-checks which presents the ESD measurements from VOICE are reliable for further VOICE galaxy-galaxy studies. The left panel of Fig.~\ref{fig:b1} shows the comparison for the $\Delta\Sigma$ of VOICE and DES lens that both have good agreements even if the ESDs of DES are more noisy than VOICE. We consider the possible reason is that the galaxies in DES survey are considered as stars or other contamination in VOICE survey because of different observational conditions of two survey projects. The fact is that VOICE has better seeing, more exposure time and more observational bands, and so on. 
    
Then we match the galaxies of shear catalogues of VOICE with DES within $1$ arcsec separation. We further measure the ESDs around FLS with MBS from matched shear catalogues of VOICE and DES, respectively. The comparison of both is displayed in the right panel of Fig.~\ref{fig:b1}. The $\Delta\Sigma$ signals from DES are some higher, and the dex of VOICE relative to DES ($\Delta\Sigma_{VOICE}-\Delta\Sigma_{DES}$) presents there are very small lower shifts. We consider the mainly reason for the offset between the ESDs of two surveys is that the shear catalogues of VOICE and DES are derived from different pipelines which are $\verb|LensFit|$ and $\verb|METACALIBRATION|$, respectively. The offset is very weak which can be ignored, and it tells the shear measurement from $\verb|LensFit|$ has good agreement with $\verb|METACALIBRATION|$. On the whole, the ESD measurements from VOICE and DES are consistent, and our results of VOICE deep survey study are reliable.

\section{The effect of redshift uncertainty}
\label{app:C}
\begin{figure}
	\centering
		\resizebox{\hsize}{!}{\includegraphics{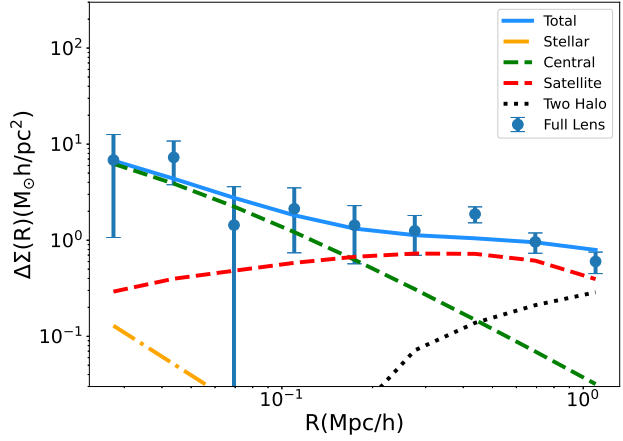}}
		\caption{ESD signals $\Delta\Sigma (P(z>z_l+0.2)>0.98$) around FLS galaxies (blue points), and the best-fitting curve (blue solid line) is comprised of the contribution from different components, which contain stellar term (orange line), central term (green line), satellite term (red line), and two halo term (black line). 
		}
		\label{fig:probs_mt0d98_model_lines}
\end{figure}
\begin{figure}
	\centering
		\resizebox{\hsize}{!}{\includegraphics{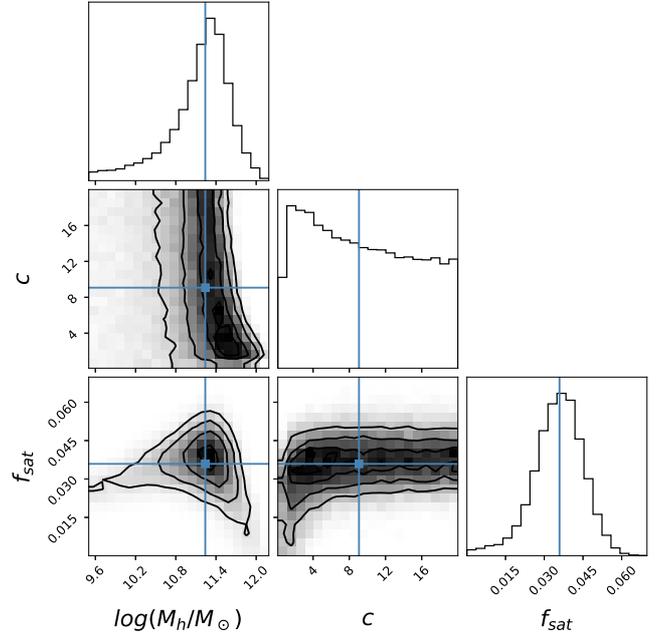}}
		\caption{Marginalised posterior distributions of three parameters obtained by using MCMC method for halo model to fit $\Delta\Sigma (P(z>z_l+0.2)>0.98$) around the galaxies of FLS.The three contour levels correspond to 16$\%$, 50$\%$, 84$\%$ confidence levels, respectively. The blue points and lines are the medians.
		}
		\label{fig:probs_mt0d98_model_constrains}
\end{figure}   
The redshift uncertainty would influence the $\Delta\Sigma$ measurements for each lens-sources pair because the surface critical density $\Sigma_{crit}$ depends on the redshift uncertainty of sources around each lens: to check this effect, we measured $\Delta\Sigma$ of FLS galaxies with the background sources selected so that the accumulated probability of photometric redshift (BPZ) $P(z_s>z_l+0.2)$ of each source is larger than 0.98 \citep{wentao2022}, 
\begin{equation}
    P(z_s>z_l+0.2)=\int^\infty_{z_l+0.2}p(z_s)dz > 0.98,
    \label{eq:probs_selection}
\end{equation}
Then the critical surface density can be calculated by
\begin{equation}
    \overline{\Sigma^{-1}_{crit}}=\frac{\int^\infty_{z_l}\Sigma^{-1}_{crit}p(z_s)dz}{\int^\infty_0p(z_s)dz}.
    \label{eq:crit_correct}
\end{equation}
We found that assuming such selection there is a negligible difference for the ESD measurement $\Delta\Sigma(P(z_s>z_l+0.2)>0.98)$ (Fig.~\ref{fig:probs_mt0d98_model_lines}), also for the model fitting curves, compared to the results displayed in Fig.~\ref{fig:sigma_fitting_lines_full}. Error bars are however larger, since there are fewer sources around each lens: this results in worse model constraints (Fig.~\ref{fig:probs_mt0d98_model_constrains}), compared to what discussed in the paper (Fig.~\ref{fig:model_constrains_full}). We therefore decided not to use the full p(z), considering that the difference is much smaller than the statistical error in our analysis.

\end{document}